\documentclass[12pt]{article}

\usepackage[top=3.5cm,bottom=3.5cm,left=2cm,right=2cm]{geometry}

\usepackage{graphicx}
\usepackage{bm}
\usepackage{amsmath,amssymb}
\usepackage{cite}
\usepackage{comment}
\usepackage{setspace}
\usepackage{float}
\usepackage{color}
\usepackage{here}

\newcommand{\nt}{\nonumber\\}

\newcommand{\ba}{\begin{eqnarray}}
\newcommand{\ea}{\end{eqnarray}}

\newcommand{\eps}{\epsilon}

\newcommand{\mathsym}[1]{{}}
\newcommand{\unicode}[1]{{}}

\begin{document}

\title{
\vbox{
\baselineskip 14pt
\hfill \hbox{\normalsize KEK-TH-2029
}} \vskip 1cm
Scale-invariant
Feature Extraction of Neural Network \\
and  
Renormalization Group Flow 
}


\author{
Satoshi Iso$^{a,b}$, Shotaro Shiba$^{a}$ and Sumito Yokoo$^{a,b}$
\vspace*{3mm}\\
\normalsize
\it $^a$\,Theory Center, High Energy Accelerator Research Organization (KEK), \\
 \normalsize
\it $^b$\,Graduate University for Advanced Studies (SOKENDAI),\\
 \normalsize
\it Tsukuba, Ibaraki 305-0801, Japan \\
}

\date{}

\maketitle

\abstract{
Theoretical understanding of how deep neural network (DNN) extracts features from input images
is still unclear, but  it is widely believed that the 
extraction is performed hierarchically through a process of coarse-graining. 
It reminds us of the basic concept of renormalization group (RG) in statistical physics. 
In order to explore possible relations between DNN and RG, 
we use the Restricted Boltzmann machine (RBM) applied to Ising model and 
construct a flow of model parameters (in particular, temperature) 
generated by  the RBM. We show that  
the unsupervised RBM trained by spin configurations at various temperatures from $T=0$ to $T=6$
generates a flow along which the temperature approaches  the critical value 
$T_c=2.2{7}$.
This behavior is opposite to the typical RG flow of the Ising model. 
By {analyzing various properties} 
 of the weight matrices of the trained RBM,  
we discuss why it flows towards $T_c$ and how 
the RBM learns to extract features of spin configurations. 
}


\newpage
\section{Introduction}

Machine learning has attracted interdisciplinary interests 
as the core method of artificial intelligence, particularly of big data science, 
and is now widely used  to discriminate subtle images 
by extracting specific features hidden in complicated input data. 
A deep neural network (DNN), which is  motivated by human brains, 
is one of  well-known algorithms \cite{Hinton1}. 
Despite its enormous successes, it is still unclear why DNN works so well 
and how DNN can efficiently extract specific features. 
In  discriminating images, we first provide samples of input images with assigned labels, 
such as a cat or a dog, and then train the neural network (NN) so as to correctly predict the labels of new, previously unseen, input images: 
this is the supervised learning and its ability of  prediction
depends on how much relevant features the NN can extract. 
On the other hand, in unsupervised learning algorithms, a NN is trained 
without assigning labels to data, 
but trained so as to generate output images that are as close 
to the input ones as possible. 
If the NN is successfully trained to reconstruct the input data, it must have acquired 
specific features of the input data. 
With this in mind, unsupervised learnings are often adopted for pre-training of supervised NNs. 

How can DNN efficiently extract features?  
Specific features characteristic to input data
usually have hierarchical structures.
An image of a cat can  still be identified as an animal  in a very low resolution image but 
one 
may not be able to distinguish it from a dog. 
Thus it is plausible that 
depth of neural networks reflects such hierarchy of features.
Namely DNN learns low-level (microscopic) characteristics 
in upper stream of the network 
and gradually extracts higher-level (macroscopic) characteristics 
as the input data flow downstream.
In other words, the initial data will get coarse-grained towards output. 
This viewpoint is reminiscent of the renormalization group (RG)  in statistical physics and 
quantum field theories, and various thoughts and studies are given 
\cite{1301.3124, Saremi, 1410.3831, 1412.6621, 1608.08225, 1609.02981, 1701.00246, 1704.06279} based on this analogy.
Especially, in a seminal paper~\cite{1410.3831}, Mehta and Schwab proposed an explicit mapping between the RG and the Restricted Boltzmann Machine (RBM) \cite{Hinton1, Hinton2, Larochelle, Hinton5, Hinton7, 1704.08724}. 

RG is the most important concept and technology to understand the critical phenomena in statistical
physics and also plays an essential role to constructively define quantum field theories on lattice. 
It is based on the idea (and proved by Kenneth Wilson \cite{Wilson}) 
 that the long-distant macroscopic  behavior of a many body system is universally 
 described by relevant operators ({\it relevant information})  around a fixed point, 
 and not affected by microscopic details in the continuum limit. 
Through reduction of degrees of freedom in RG, the relevant information is 
{emphasized} 
while other irrelevant information is discarded. 
Particularly, suppose that the statistical model is described by a set of parameters 
$\{ \lambda_\alpha \}$, and that  the parameters are mapped to a different set
$\{ \tilde\lambda_\alpha \}$ by RG transformations\footnote{\label{RGparameters} In order to describe
the RG transformation exactly, infinitely many parameters are necessary to be introduced.  But
it can be usually well-approximated by a finite number of parameters.}. 
Repeating such RG transformations,
we can draw a flow diagram in the parameter space of the statistical model,
\begin{equation}
 \{ \lambda_\alpha \} \rightarrow \{ \tilde \lambda_\alpha \} \rightarrow \{ \tilde{\tilde{\lambda}}_\alpha \} 
 \rightarrow \cdots .
 \label{parameter-flow}
\end{equation}
These RG flows control the behavior of the statistical model  near the critical point where
a second order phase transition occurs. 

A simplest version of RBM is a NN consisting of two layers, 
a visible layer with variables $\{v_i =\pm 1 \}$
and a hidden layer with variables $\{ h_a =\pm 1\}$,  
that are coupled to each other through the Hamiltonian
\begin{equation}
\Phi(\{v_i\}, \{h_a \})=-(\sum_{i,a} W_{ia}v_i h_a +\sum_i b_i^{(v)} v_i + \sum_a b_a^{(h)} h_a ) \,.
\label{RBM-Hamiltonian}
\end{equation} 
A probability distribution of a configuration $\{v_i ,h_a  \}$ is given by
\begin{equation}
p(\{v_i \},\{h_a  \}) = \frac{1}{\cal Z} e^{-\Phi(\{v_i\}, \{h_a \})}
\label{prob-vh}
\end{equation}
where we defined the partition function by
${\cal Z} = \sum_{\{v_i, h_a\}} e^{-\Phi(\{v_i\}, \{h_a \})}$.
No intra-layer couplings are introduced in the RBM. 
Now  suppose that the RBM is already trained and the parameters of the Hamiltonian 
(\ref{RBM-Hamiltonian}), namely $\{ W_{ia}, b_i^{(v)}, b_a^{(h)} \}$,  are already fixed
through a process of training. 
The probability distribution $p(\{v_i \},\{h_a  \})$ also provides the following conditional probabilities for 
$\{ h_a \}$ (or $\{v_i \})$  with the other variables being kept fixed;
\ba 
&& p(\{h_a\}| \{ v_i \}) =  \frac{p(\{h_a\}, \{ v_i \})}{\sum_{\{h_a\}}  p(\{h_a\}, \{ v_i \}) } 
 \label{cond-p-h}  \\
&&    p(\{ v_i\}| \{ h_a \}) = \frac{p(\{h_a\}, \{ v_i \})}{\sum_{\{v_i\}}  p(\{h_a\}, \{ v_i \}) } \,.
\label{cond-p-v}
\ea
These conditional probabilities generate a flow of distributions, and consequently a flow 
of parameters $\{\lambda_\alpha \}$ of the corresponding statistical model.
Suppose that we have a set of  $N (\gg 1)$ initial configurations
$\{ v_i=\sigma_i^{A} \}$  ($A=1, \ldots, N$), which are generated  by 
a statistical model with parameters $\lambda_\alpha$, such as the Ising model at temperature $T$. 
In the large $N$ limit, the distribution function 
  \begin{equation}
 q_0(\{v_i\})= \frac{1}{N} \sum_A \delta (v_i - \sigma_i^{A})
 \label{q_data}
 \end{equation}
faithfully characterizes the statistical model with parameters $\lambda_\alpha$.
Multiplying $q_0(\{v_i \})$ by the conditional probabilities (\ref{cond-p-h}) and (\ref{cond-p-v}) iteratively,
we can generate a flow of probability distributions as
\ba
&& q_0(\{v_i\}) \rightarrow r_1(\{h_a \})= \sum_{\{v_i\}} p(\{h_a\}| \{ v_i \}) q_0(\{v_i\})  \label{q0r1} 
\label{prob-flows1}
\\
&& r_1(\{h_a \}) \rightarrow q_1(\{v_i\}) = \sum_{\{h_a\}} p( \{ v_i \}| \{h_a\}) r_1(\{h_a\})
\label{prob-flows}
\ea
and so on for $q_n(\{v_i\}) \rightarrow r_{n+1}(\{h_a \})$ and $r_{n+1}(\{h_a \}) \rightarrow q_{n+1}(\{v_i\})$.
Let us focus on Eq.\,(\ref{q0r1}).
If the probability distribution $r_1(\{h_a \})$ is well approximated by the Boltzmann distribution
of the same statistical model    with different parameters  $\tilde\lambda_{\alpha}$,
we can say that the RBM generates 
a transformation\footnote{The situation is similar to the footnote \ref{RGparameters},
and infinitely many parameters are necessary to represent the probability distribution $p(\{h_a\})$
in terms of the statistical model.}
 from $\{ \lambda_\alpha \}$ to $\{ \tilde\lambda_{\alpha} \}$. 
If more than two layers are 
stacked iteratively, 
we can obtain a flow of parameters as in Eq.\,(\ref{parameter-flow}). 
Another way to obtain a flow is to look at
the transformations 
$q_0(\{ v_i \}) \rightarrow q_1(\{ v_i \}) \rightarrow q_2(\{ v_i \}) \rightarrow \cdots$
and translate the flow of probability distributions  into a flow of parameters $\{ \lambda_\alpha \}.$ 
In the present paper, we consider the latter flow to discuss a relation with RG. 

 Mehta and Schwab \cite{1410.3831} pointed out  similarity between RG transformations
 of Eq.\,(\ref{parameter-flow})
 and the above flows of parameters in the unsupervised RBM. 
 But in order to show that the
 transformation of parameters $\{ \lambda_\alpha \}$
  in the RBM indeed generates the conventional RG transformation, 
 it is necessary to show that 
the weight matrix $W_{ia}$ and the biases  $b_i^{(v)}, b_a^{(h)}$
 of the RBM
 must be appropriately chosen so as to generate the correct RG transformation 
 that performs coarse-graining of input configurations.
 In Ref.\cite{1410.3831}, multi-layer RBM is employed as an unsupervised-learning NN,
 and the weights and the biases are chosen by minimizing the KL divergences (relative entropy) between 
 the input probability distribution and the reconstructed distribution 
 by  integrating (marginalizing) over the hidden variables.
 The authors suggested the similarity by looking at the local spin structures in the hidden variables, 
 but they did not show it explicitly that the weights determined 
 by the unsupervised learning actually generate the flow of RG transformations.

The arguments \cite{1410.3831}  and misconception  in the literature are criticized by
Ref.\cite{1608.08225}.
 In a wider context, the criticism is related to the following question: 
 what determines whether 
 a specific feature of input data is relevant or not? In RG transformations of statistical models, 
 long-wave length (macroscopic) modes are highly respected
while 
short-wave length modes are discarded as noise. 
In this way, RG transformations can extract universal behavior of the model 
 at long-wave length. But, of course,  it is so  because we are interested in the macroscopic 
behavior of the system:  if we are instead interested in short-wave length physics,
we need to extract opposite features of the model.  Thus, we may say that
extraction of  {\it relevant features} needs pre-existing biases to judge,
and supervised learning is necessary to give such biases to the machine.
However  
this does not mean that unsupervised learnings do not have anything to do with the RG.
Even in unsupervised learnings, a NN automatically notices and
 extracts some kind of features of  the input data and 
the flow generated by the trained NN reflects such features. 

In the present paper, we investigate relationship between the RBM and the RG by further
studying the flows of distributions, Eqs.\,(\ref{prob-flows1}) and (\ref{prob-flows}),  
that the unsupervised RBM generates. 
Here notice that in defining the flow of (\ref{prob-flows1}) and (\ref{prob-flows}), we need to specify  
how we have trained the RBM because the training determines 
the properties of the weights and biases, and accordingly the behavior of the flow.
In this paper we mostly use the following three different ways of trainings. 
One type of RBM (we call type V) is trained by configurations at various temperatures from low to high. 
Other two types (type H and L) are trained by configurations only at high (and only at low) temperatures. 
Then we translate these flows of probability distributions 
defined by Eqs.\,(7) and (8)
into flows of  temperature of the Ising model,
\begin{equation}
  T \rightarrow \tilde{T}  \rightarrow \tilde{\tilde{T}}   \rightarrow \cdots .
  \label{Ising-flow}
\end{equation}
In order to measure temperature, 
we prepare another NN trained by a supervised learning. 
The results of our numerical simulations  lead to  a surprising conclusion.   
In the type V RBM {that has adequately learned  the features of configurations at various temperatures,}
we found that
 the temperature {\it approaches the critical point}, $T \rightarrow T_c$, along the RBM flow.
The behavior is opposite to the conventional RG flow of Ising model.  


The paper is organized as follows.
In section \ref{methods}, we explain the basic settings and the methods of our investigations.
We prepare sample images of the spin configurations of the Ising model,
and train  RBMs by the configurations without assigning labels of temperature.
Then we construct flows of parameters (i.e., temperature) generated by the trained 
RBM\footnote{
Two-dimensional Ising model is the simplest statistical model to exhibit the second order phase transition,
and there are many previous studies of the Ising model using machine learnings.
See e.g., \cite{1606.00318, 1606.02718, Tanaka:2016rtu, 1703.02435, 1704.00080, 1708.04622}.}.
In section \ref{sec:res}, we show various results of the numerical simulations, including the RBM flows of parameters.
In section \ref{weights}, we analyze properties of the weight matrices $W_{ia}$ using the method of 
singular value decomposition.
The final section is devoted to summary and discussions. 
Our main results of the RBM flow and conjectures about the feature extractions of the unsupervised RBM
 are written in Sec.\,\ref{sec:res_rbm}.

\section{Methods}
\label{methods}

We  explain various methods for numerical simulations to investigate relations between 
the unsupervised RBM and the RG  of   Ising model. 
Though most methods in this section are standard {and well known}, we explain them in details 
to make the paper self-contained. 
In Sec.\,\ref{method-Flow},  {we explain the central method of 
generating the RBM flows.}
 Basic materials of the RBM are given in Sec.\,\ref{method-USRBM}. 
The other two sections, Secs.\,\ref{MCsimulation} and \ref{sec:NN_set}, can be skipped over unless one is interested in
how we generate the initial spin configurations and  measure  temperature of a set of configurations.


\subsection{Monte-Carlo simulations of Ising model}
\label{MCsimulation}

We first construct samples of configurations of the 
two-dimensional Ising model by using Monte-Carlo simulations. 
The spin variables $\sigma_{x,y} =\pm 1$ are defined on a two dimensional lattice 
of size  $L\times L$. The index $(x,y)$ represents each lattice site 
and takes
$x,y = 0,1,\ldots,L-1$. 
The Ising model Hamiltonian is given by
\ba
H = -J \sum_{x,y=0}^{L-1} \sigma_{x,y} \left( \sigma_{x+1,y} + \sigma_{x-1,y} + \sigma_{x,y+1} + \sigma_{x,y-1} \right). 
\ea
It describes a ferromagnetic model for $J>0$ and an anti-ferromagnetic model for $J<0$.
Here we impose the periodic boundary conditions for the spin variables,
\ba
\sigma_{L,y} := \sigma_{0,y} \,,\quad
\sigma_{-1,y} := \sigma_{L-1,y} \,,\quad
\sigma_{x,L} := \sigma_{x,0} \,,\quad
\sigma_{x,-1} := \sigma_{x,L-1}\,.
\ea

Generations of spin configurations  at temperature $T$ are performed by
the method of Metropolis Monte Carlo (MMC) simulation. 
In the method, we first generate a random configuration $\{\sigma_{x,y}\}$.
We then  choose one of the spins $\sigma_{x,y}$ and
flip its spin with the probability
\ba
p_{x,y} = \begin{cases}
1 & (\text{when}~dE_{x,y}<0) \\
e^{-dE_{x,y} / k_B T} & (\text{when}~dE_{x,y}>0)\,
\label{flip-prob}
\end{cases}
\ea
where $dE_{x,y}$ is the change of energy of this system
\ba
dE_{x,y} = 2J \sigma_{x,y} \left( \sigma_{x+1,y} + \sigma_{x-1,y} + \sigma_{x,y+1} + \sigma_{x,y-1} \right).
\ea
The probability of flipping the spin (\ref{flip-prob}) satisfies the detailed balance condition
$P_{s \rightarrow s'} \rho_s = P_{s' \rightarrow s} \rho_{s'}$ where $\rho_s \propto e^{-E_s/k_B T}$
is the canonical distribution of the spin configuration $s=\{\sigma_{x,y} \}$ at temperature $T$.
Thus after many iterations of flipping all the spins, the configuration approaches the equilibrium distribution at $T$.  
Since all physical quantities are written in terms of a combination of $J/k_B T$,
we can set the Boltzmann constant $k_B$ and the interaction parameter $J$ to be equal to 1
without loss of generality.

\begin{figure}[h]
\begin{center}
\includegraphics[width=10.6cm, height=3cm]{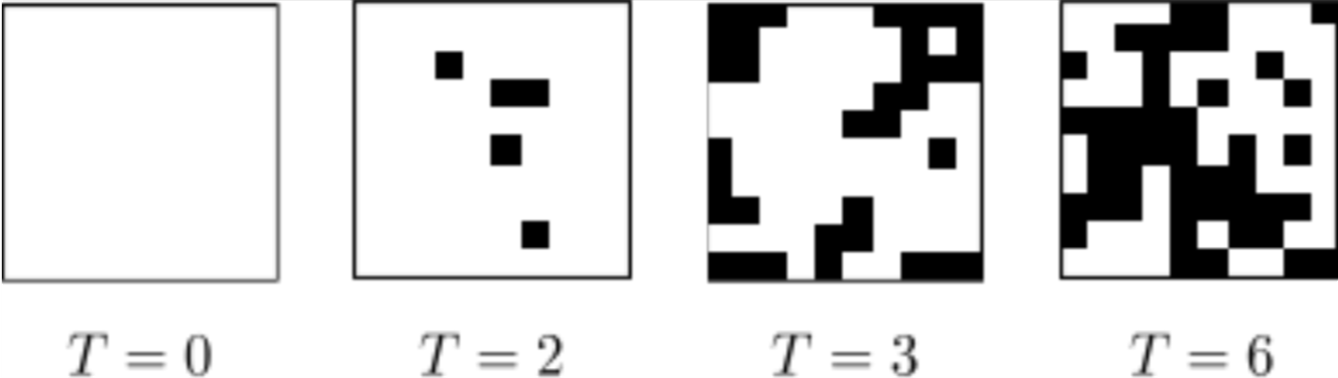}
\caption{\small Examples of spin configurations at temperatures $T=0,2,3,6$}
\label{config}
\end{center}
\end{figure}

In the following analysis, we set the lattice size $L^2=10 \times 10$ 
and repeat the procedure of MMC simulations $100L^2 = 10000$ times to construct spin configurations. 
In our simulations,
we generated spin configurations at various temperatures  
$T=0, 0.25, 0.5,\ldots, 6$.\footnote{
For $T=0$, we practically set $T=10^{-6}$ for numerical calculations.}
{Some of typical spin configurations are shown in Fig.\,\ref{config}.}

\subsection{Unsupervised learning of the RBM} 
\label{method-USRBM}

Our main motivation in the present paper is to study whether 
the RBM is related to the RG in statistical physics. 
In this section, we {review the basic algorithm} 
 of the RBM \cite{Hinton1, Hinton2, Larochelle, Hinton5, Hinton7, 1704.08724}
which is  trained by 
the configurations constructed by the MMC method of Sec.\,\ref{MCsimulation}.
\begin{figure}[h]
\begin{center}
\includegraphics[width=8.5cm, height=5.8cm]{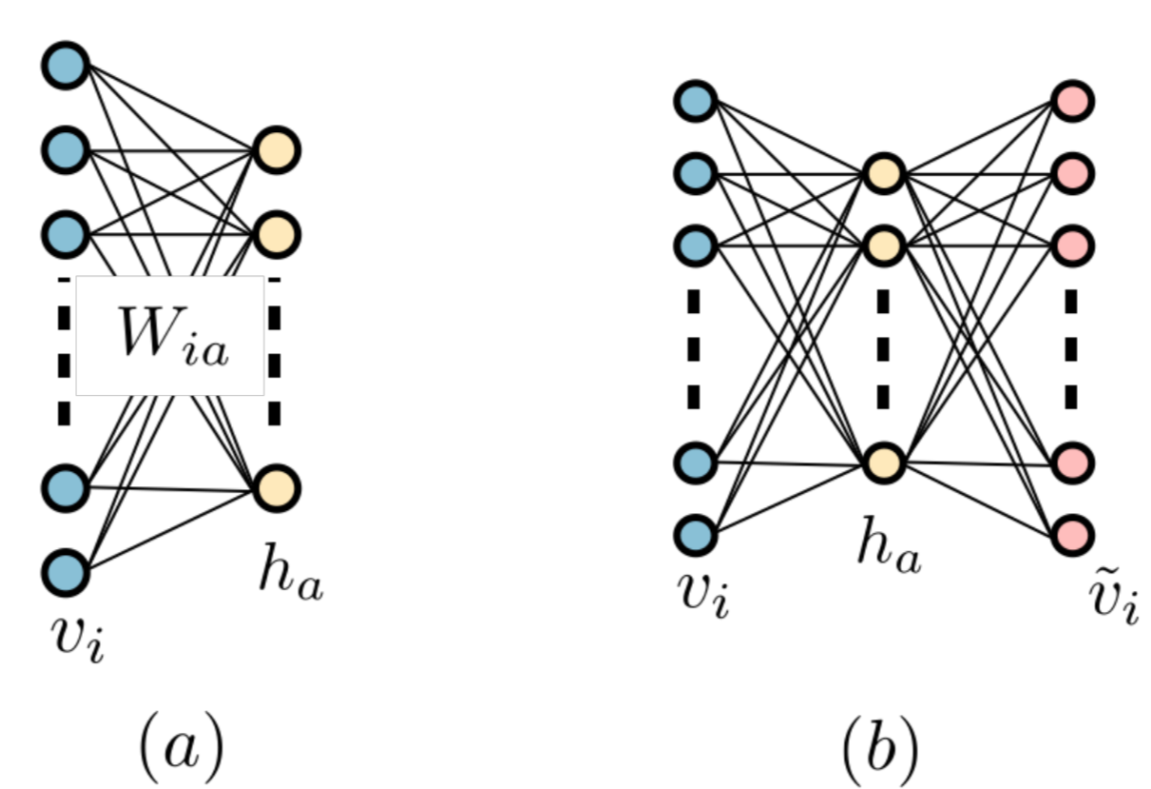}
\caption{\small (a) Two-layer neural network of the RBM with a visible layer $\{v_i\}$ and a hidden layer $\{h_a\}$. These two layers are coupled but there are no intra-layer couplings.
(b) The RBM generates reconstructed configurations from $\{v_i\}$ to $\{\tilde{v}_i\}$ through the hidden configuration $\{h_a\}$.
}
\label{figRBM}
\end{center}
\end{figure}
As explained in the Introduction, the RBM consists of two layers as shown {in the left panel of} Fig.\,\ref{figRBM}.
The initial configurations {$\{\sigma_{x,y}\}$} of Ising model generated
 at various temperatures are input into 
the visible layer $\{v_i\}$. The number of neurons in the visible layer is fixed at  
$N_v=L^2=100$ ($i= 1, \ldots, N_v$) to represent 
the spin configurations of Ising model. 
On the other hand, {the  hidden layer can take an arbitrary number of neurons, $N_h$. 
In the present paper, we consider 7 different sizes;}
$N_h= 16, 36, 64, 81, 100, 225$ and 400. 
Thus the $N_h$ spin variables in the hidden layer are given by $\{h_a\}$ for $ a=1, \ldots, N_h$.

The RBM is a generative model of probability distributions based on Eq.\,(\ref{prob-vh}).
We first explain how we can train the RBM by optimizing 
the weights $W_{ia}$ and the biases $b_i^{(v)}, b_a^{(h)}$.
Our goal is
to represent the given probability distribution $q_0(\{v_i\})$ in Eq.\,(\ref{q_data}),
as faithfully as possible, in terms of a model probability distribution defined by
\begin{equation}
p(\{v_i \}) = \frac{1}{\cal Z}  \sum_{\{h_a\}} e^{-\Phi(\{v_i\}, \{h_a \})} .
\end{equation} 
The partition function ${\cal Z}=\sum_{\{v_i,h_a\}} e^{-\Phi(\{v_i\}, \{h_a \})}$ is difficult to evaluate, 
but summations over only 
one set of spin variables (e.g. over $\{v_i\}$) are easy to perform because of the absence 
of the intra-layer couplings. It also makes the conditional probabilities 
(\ref{cond-p-h}) and (\ref{cond-p-v}) to be rewritten as products of probability distributions
of each spin variable;
\ba
&& p(\{h_a\}| \{v_i\}) = \prod_a p(h_a| \{v_i \}) = \prod_a
\frac{1}{1 + \exp\left[- 2 h_a \left(\sum_i W_{ia} v_i + b^{(h)}_a \right)\right]} 
\label{p_model} \\
&&  p(\{ v_i\}| \{ h_a \}) = \prod_i  p( v_i | \{ h_a \}) 
= \prod_i \frac{1}{1 + \exp\left[- 2 {v}_i \left(\sum_a W_{ia} h_a + b^{(v)}_i \right)\right]}  \,.
\label{p_tilde}
\ea
Then the expectation values of spin variables in the hidden (or visible) layer
 in the background of spin configurations in the other layer are calculated 
 as
\ba
&& \langle h_a \rangle_{\{v_i\}} = \tanh \left( \sum_i W_{ia} v_i + b^{(h)}_a \right) 
\label{vev-h-inv} \\
&& \langle  v_i \rangle_{ \{h_a \} }  = \tanh\left(\sum_a W_{ia} h_a + b^{(v)}_i \right) .
\label{vev-v-inh}
\ea


Now the task is to train the RBM so as to minimize 
the distance between two probability distributions of $q(\{v_i \})$ and $p(\{v_i \})$
by appropriately choosing the weights and the biases.
The distance is called 
Kullback-Leibler (KL) divergence, or relative entropy, and given by
\ba \label{KL}
\text{KL}(q|| p) = \sum_{\{v_i\}} q(\{v_i\}) \log \frac{q(\{v_i\})}{ p(\{v_i \})} 
= \text{const.} -  \sum_{\{ v_i \}} q(\{v_i\}) \log p(\{v_i \})\,.
 \ea
If two probabilities are equal, the KL divergence vanishes. 
Otherwise derivatives of KL$(q||p)$ with respect to the weight $W_{ia}$ and the biases 
$b_i^{(v)}, b_a^{(h)}$  are given by 
\ba
\frac{\partial \text{\,KL}(q|| p) }{\partial W_{ia}} &=& \langle  v_i  h_a  \rangle_{\text{data}} 
- \langle  v_i h_a\rangle_{\text{model}}  \nonumber \\
\frac{\partial \text{\,KL}(q|| p) }{\partial b_{i}^{(v)}} &=& \langle  v_i \rangle_{\text{data}} 
- \langle  v_i \rangle_{\text{model}} \label{KL-full}  \\
\frac{\partial \text{\,KL}(q|| p) }{\partial b_{a}^{(h)}} &=& \langle  h_a \rangle_{\text{data}} 
- \langle  h_a \rangle_{\text{model}}\,, \nonumber
\ea
where averages are defined by
\ba
&& \langle  A(\{ v_i \}) \rangle_{\text{data}} = \sum_{\{ v_i \} } q(\{v_i \}) A(\{ v_i \}) \\
&& \langle  A(\{ v_i \}, \{h_a \}) \rangle_{\text{model}} = \sum_{\{ v_i \}, \{h_a\} } p(\{v_i \}, \{h_a\}) A(\{ v_i \}, \{h_a \}) \,,
\ea
and  $h_a$ in $\langle \cdots  \rangle_{\text{data}}$ is replaced by 
$ \langle h_a \rangle_{\{v_i\}} $ of Eq.\,(\ref{vev-h-inv}).
In training the RBM,
we change {the weights and biases} so that the KL divergence is 
reduced. 
Using the method of back propagation~\cite{Amari},
we renew values of the weights and biases as
\ba
W ~&\to&~ W_\text{new} = W + \delta W \nt
b^{(v)} ~&\to&~ b^{(v)}_\text{new} = b^{(v)} + \delta b^{(v)} \nt
b^{(h)} ~&\to&~ b^{(h)}_\text{new} = b^{(h)} + \delta b^{(h)}
\label{renewalRBM}
\ea
where 
\ba
\delta W_{ia} &=& \eps \left( \langle v_i h_a\rangle_{data} - \langle v_i h_a \rangle_{model} \right) \nt
\delta b^{(v)}_i &=& \eps \left( \langle v_i \rangle_{data} - \langle v_i \rangle_{model} \right) \nt
\delta b^{(h)}_a &=& \eps \left( \langle h_a\rangle_{data} - \langle h_a \rangle_{model} \right)\,.
\ea
Here $\eps$ denotes the learning rate, which we set to 0.1.
The first terms $\langle \cdots \rangle_{{data}}$ are easy to calculate, but 
the second terms $\langle \cdots \rangle_{{model}}$
are difficult to evaluate since it requires the knowledge of the full partition function ${\cal Z}$.

To avoid this difficulty, 
we {need to} use the method of Gibbs sampling 
to approximately evaluate these expectation values $\langle \cdots \rangle_{{model}}$.
{Practically we employ a more simplified method,} 
which is called  the method of contrastive divergence (CD) \cite{Hinton4, Hinton8, Bengio}. 
The idea is very simple, and reminiscent of the mean field approximation in statistical physics. 
Given the input data of the visible spin configurations $\{ v_i^{A(0)}=\sigma_i^{A} \}$,
the expectation value of the hidden spin variable $h_a$ 
 can be easily calculated as Eq.\,(\ref{vev-h-inv}). We write the expectation value as
\ba
h_a^{A(1)} := \langle h_a \rangle_{\{v_i^{A(0)}\}} = \tanh \left( \sum_i W_{ia} v_i^{A(0)} + b^{(h)}_a \right) .
\label{v0-h1}
\ea
Then in this background of the hidden spin configurations, the expectation value of $v_i$ can be
again easily calculated by using Eq.\,(\ref{vev-v-inh}). We write it as
\ba
v_i^{A(1)} :=  \langle  v_i \rangle_{ \{h_a^{A(1)} \} }  = \tanh\left(\sum_a W_{ia} h_a^{A(1)} + b^{(v)}_i \right) .
\label{h1-v1}
\ea
Then we obtain $h_a^{A(2)}= \langle h_a \rangle_{\{v_i^{A(1)}\}}$, and so on.  
We can iterate these procedure many times and replace the second terms in Eq.\,(\ref{KL-full}) by
the expectation values generated by this method. 

In doing the numerical simulations in the present paper, we adopt 
the simplest version of CD, called CD${}_1$, which gives us
the following approximate formulas:
\ba
\langle v_i \rangle_{data} =
 \frac{1}{N} \sum_A \sigma_{i}^{A} ,\quad
\langle h_a \rangle_{data}
= \frac{1}{N} \sum_A h_a^{A(1)} ,\quad
\langle v_i h_a \rangle_{data} = 
 \frac{1}{N} \sum_A \sigma_{i}^{A}  h_a^{A(1)} ,
\ea
and 
\ba
\langle v_i \rangle_{model}
= \frac{1}{N} \sum_A  v_i^{A(1)} ,\quad
\langle h_a \rangle_{model}
= \frac{1}{N} \sum_A  h_a^{A(2)} ,\quad
\langle v_i h_a \rangle_{model} =
 \frac{1}{N} \sum_A  v_i^{A(1)} h_a^{A(2)} .
\ea
Here $\sigma_i^{A}$ denotes each spin configuration $\{\sigma_{x,y}\}$ generated 
by the method of Sec.\,\ref{MCsimulation}.
As input data to train the RBM, 
we generated 1000 spin configurations for  each of 25 different temperatures $T=0,0.25,\ldots,6$.
Then 
the index $A$ runs from $1$ to $N=25000$.
In some cases, as we will see in Sec.\,\ref{sec:res_rbm}, we use only 
a restricted set of configurations at high or low temperatures,
then 
the index runs $A=1,\ldots,N=1000\times$(number of temperatures).

We repeat the renewal procedure (\ref{renewalRBM}) many times 
(5000 epochs), 
and obtain adjusted values of the weights and biases.
In this way we train the RBM by using a set of
configurations $\{ v_i^{A(0)}=\sigma_i^{A} \}$, $(A=1, \ldots, N)$.

\subsection{Generation of RBM flows}
\label{method-Flow}

As discussed in the Introduction, once the RBM is trained and the weights and biases are fixed,
the RBM generates a sequence of probability distributions (\ref{prob-flows}).
Then we  translate the sequence into a flow of parameters (i.e., temperature). 
In generating the sequence, the initial set of configurations 
should be prepared  {separately in addition to} 
the  configurations that are used to  train the RBM\footnote{Thus we  generate 
$25000$ spin configurations {in addition to} the {$25000$} 
configurations used for training the RBM.}.

We can also generate a flow of parameters in a slightly different way.
For a specific configuration $v_i=v_i^{(0)}$, we can define a sequence of configurations
following Eqs.\,(\ref{v0-h1}) and (\ref{h1-v1}) as
\ba
\{v_i^{(0)}\} \rightarrow \{h_a^{(1)}\} \rightarrow \{v_i^{(1)}\} \rightarrow \{h_a^{(2)}\}
 \rightarrow \{v_i^{(2)} \}\rightarrow  \cdots\,.
\ea
{The right panel of Fig.\,\ref{figRBM} shows a generation of new configurations from
$\{v_i\}$ to $\{\tilde{v}_i \}$ through $\{ h_a\}$.}
Since each value of $v_i^{(n)}$ and $h_a^{(n)}$ (for $n>0$) is defined by 
an expectation value as in Eqs.\,(\ref{v0-h1}) and (\ref{h1-v1}), 
it does not take an integer $\pm 1$ but a fractional value between  $\pm 1$. 
In order to get a flow of spin configurations, we need to
replace these fractional values by $\pm 1$ 
with a probability $ (1\pm \langle v_i^{(n)}\rangle)/2$  {or $(1 \pm \langle h_a^{(n)} \rangle)/2$.}
It turns out that the replacement is usually a good approximation since 
the expectation values are likely to take values  close to $\pm 1$ owing to the property of the trained weights
 $|W_{ia}| \gg 1$. In this way, we obtain a flow of spin configurations
\ba
\{v_i^{(0)}\}  \rightarrow \{v_i^{(1)}\}  \rightarrow \{v_i^{(2)} \} \rightarrow
\cdots  \rightarrow \{v_i^{(n)}\} 
\label{config-flow-v}
\ea 
starting from  the initial configuration $\{v_i^{(0)}\}.$
The flow of configurations is transformed to a flow of temperature  distributions
by using the method explained in Sec.\,\ref{sec:NN_set}. 

\subsection{Temperature measurement by a supervised-learning NN}
\label{sec:NN_set}

\begin{figure}[h]
\begin{center}
\includegraphics[width=4.7cm, height=6cm]{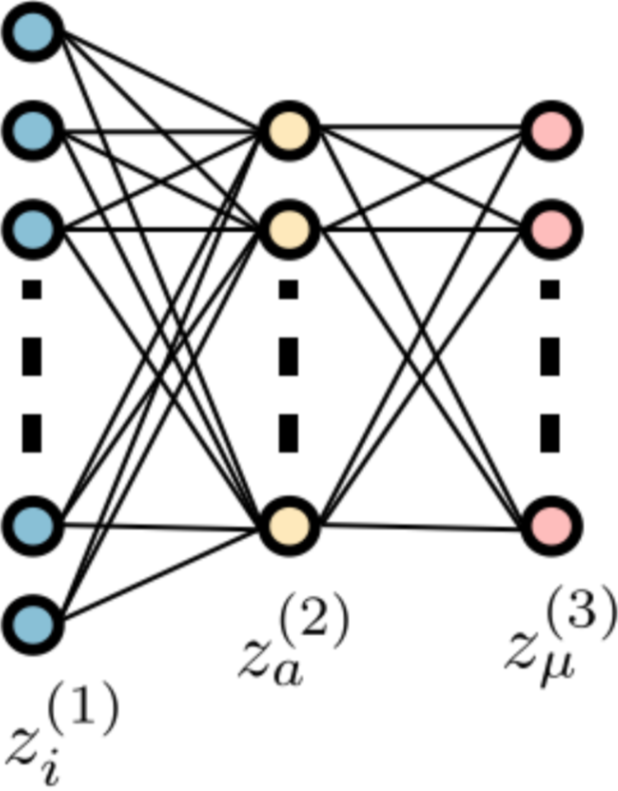}
\caption{\small Three-layer neural network for supervised learning with an input layer $\{z_i^{(1)}\}$, a hidden layer $\{z_a^{(2)}\}$ and an output layer $\{z_\mu^{(3)}\}$.}
\label{figSNN}
\end{center}
\end{figure}

Next we design a neural network (NN) to measure temperature of  spin configurations.
The NN for the supervised learning has three layers with one hidden layer in the middle
(See  Fig.\,\ref{figSNN}).

The input layer $\{z^{(1)}_i\}$ consists of $L^2=100$ neurons in which we input 
 spin configurations of Ising model.
The output layer $\{z^{(3)}_\mu\}$ has $25$ neurons which
 correspond to 25 different temperatures that we want to measure.
The number of neurons in the hidden layer $\{z^{(2)}_a \}$ is set to $64$.
We train this three-layer NN by a set of spin configurations, each of which has a label of temperature.
Thus this is  the supervised learning. 
As  input data to train the NN, 
{we use the same $N=25000$ configurations which were used to train the RBM}\footnote{
In order to check the performance of the NN, namely to see how precisely the machine can measure 
the temperature of a new set of configurations,  
we use  other 25000 configurations that are prepared for generating 
the sequence of probability distributions of the RBM  in Sec.\,\ref{method-Flow}.
We will show the results of the performance in Sec.\,\ref{sec:res_sl}.
}.


The training of the NN is carried out as follows.
Denote the input data as
\ba
Z^{(1)}_{Ai} = \sigma_{i}^{A}
\ea
where $A=1,\ldots,N$, and $\sigma_{i}^{A}$ are the spin configurations $\{\sigma_{x,y}=\pm 1\}$ as in Sec.\,\ref{method-USRBM}. 
The input data is transformed to $Z^{(2)}_{Aa}$ in the hidden layer 
by the following nonlinear transformation;
\ba
Z^{(2)}_{Aa} = f\left( \sum_{i=1}^{100} Z^{(1)}_{Ai} W^{(1)}_{ia} + 
b^{(1)}_a \right) 
=: f\left(U^{(1)}_{Aa}\right)
\ea
where $W^{(1)}_{ia}$ is an weight matrix and $b^{(1)}_a$ is a bias.
The activation function $f(x)$ is chosen as $f(x)=\tanh(x)$.
$Z^{(2)}_{Aa}$ is transformed to $Z^{(3)}_{A\mu}$ in the output layer, which corresponds 
to the label, namely temperature, of each configuration.
The output $Z^{(3)}_{A\mu}$ is given by
\ba
Z^{(3)}_{A\mu} = g\left(\sum_{a=1}^{64} Z^{(2)}_{Aa} W^{(2)}_{a\mu} + 
b^{(2)}_\mu \right)
=: g\left(U^{(2)}_{A\mu}\right)
\ea
where $W^{(2)}_{a\mu}$ and $b^{(2)}_\mu$ are another weight matrix and bias.
The function $g(x)$ is the softmax function
\ba
g\left(U^{(2)}_{A\mu}\right) = \frac{\exp U^{(2)}_{A\mu}}{\sum_{\nu=1}^{25} \exp U^{(2)}_{A\nu}}\,,
\ea
so that $Z^{(3)}_{A\mu}$ can be regarded as a probability 
since $\sum_\mu Z^{(3)}_{A\mu}=1$ is satisfied for each configuration $A$. 
Thus the NN transforms
an input spin configuration $Z^{(1)}_{Ai}$  to the probability
$Z^{(3)}_{A\mu}$ of the configuration to take the $\mu$-th output value (i.e., temperature).

Each of the input configurations $Z^{(1)}_{Ai}$ is generated by the MMC method
at temperature $T$.
$T$ takes one of the 25 discrete values 
 $T=\frac{1}{4} (\nu-1)$, ($\nu=1,\ldots,25$). 
 If the A-th configuration  is labelled by  $\nu$, 
we want  the NN  to give an output $Z_{A\mu}^{(3)}$ as close as 
the following one-hot representation:
\ba
d_{A}^{(\nu)} = (0,\ldots,0,  \check{1}^{(\nu)}, 0,\ldots,0)_A\,,
\ea
or its $\mu$-th component is given by $d_{A\mu}^{(\nu)}=\delta_{\mu\nu}$.
It can be  interpreted as a probability of the configuration $A$  to 
take the  $\mu$-th output. 
Then the task of the supervised training is to minimize the cross entropy, which is equivalent to
the KL divergence of the desired probability $d_{A\mu}^{(\nu)}$ and the output probability $Z^{(3)}_{A\mu}$.
The loss function is thus given by the cross entropy,
\ba \label{loss_sl}
E_A = \text{KL}(d_{A\mu}^{(\nu)}||Z^{(3)}_{A\mu}) =
-\sum_\mu d_{A\mu}^{(\nu)} \log Z^{(3)}_{A\mu}\,.
\ea
Then,  using the method of back propagation, we renew values of 
the weights and biases from the lower to the upper stream;
\ba
W^{(\ell)} ~&\to&~ W^{(\ell)}_\text{new} = W^{(\ell)} + \delta W^{(\ell)} \nt
b^{(\ell)} ~&\to&~ b^{(\ell)}_\text{new} = b^{(\ell)} + \delta b^{(\ell)}.
\label{WBrenewal-TM}
\ea
The variations of  $\delta W^{(\ell)}, \delta b^{(\ell)}$ at the lower stream are given by
\ba
\delta W^{(2)}_{a\mu} 
&=& -\frac{\eps}{N} \sum_A (Z^{(2)})^T_{aA} \Delta^{(3)}_{A\mu} \nt
\delta b^{(2)}_\mu
&=& -\frac{\eps}{N} \sum_A \Delta^{(3)}_{A\mu}
\ea
where $\Delta^{(3)}_{A\mu} = Z_{A\mu}^{(3)} - d_{A\mu}^{(\nu)}$. 
The learning rate $\eps$ is set to $0.1$.
Then using these lower stream variations,
we change the upper stream weights and biases as
\ba
\delta W^{(1)}_{ia} 
&=& -\frac{\eps}{N} \sum_A (Z^{(1)})^T_{iA} \Delta^{(2)}_{Aa} \nt
\delta b^{(1)}_a
&=& -\frac{\eps}{N} \sum_A \Delta^{(2)}_{Aa}
\ea
where
\ba
\Delta^{(2)}_{Aa} &=& \sum_\mu \Delta^{(3)}_{A\mu} (W^{(2)})^T_{\mu a} 
 f'\left(U^{(1)}_{Aa}\right) .
\ea
We repeat this renewal procedure many times (7500 epochs) for the training 
of the NN to
 obtain suitably adjusted values of the weights and biases.

Finally we note how we measure temperature of a configuration. 
If the size of a  configuration generated at temperature $T$ is large enough, 
say $L=10^{10^{100}}$, the 
trained NN will reproduce the temperature of the configuration quite faithfully. 
However our configurations are small sized with only $L=10$. 
Thus we instead need an ensemble of many spin configurations and measure a temperature distribution of the configurations. 
The supervised learning gives us this probability distribution of temperature. 

\section{Numerical results}
\label{sec:res}

In this section we present our numerical results for the flows generated by unsupervised RBM,
and  discuss a relation with the renormalization group flow of Ising model. 
Our main results of the RBM flows are written in Sec.\,\ref{sec:res_rbm}.

\subsection{Supervised learning for temperature measurement}
\label{sec:res_sl}

Before discussing the unsupervised RBM, 
let us first see how we trained the NN  to measure temperature. 

\begin{figure}[h]
\begin{center}
\includegraphics[width=8cm, height=6cm]{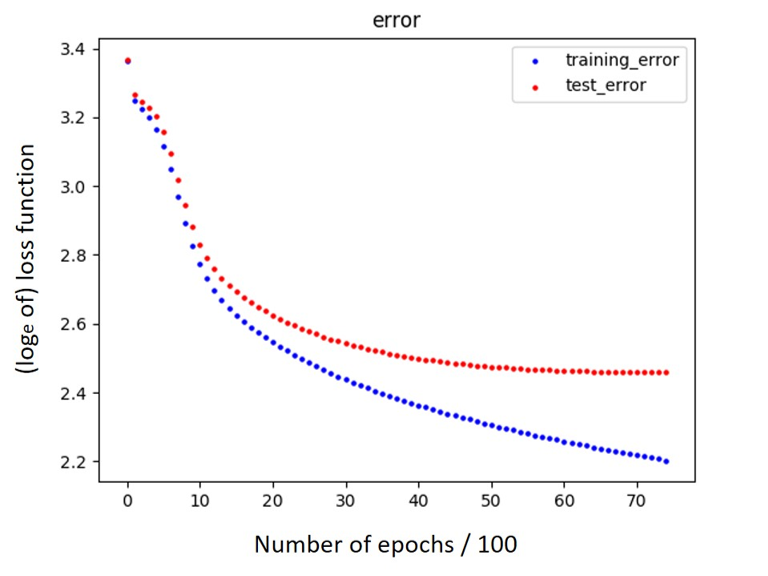}
\caption{\small Training error and test error (up to 7500 epochs)} 
\label{fig1}
\end{center}
\end{figure}

\begin{figure}[!h]
\begin{center}
\includegraphics[width=8cm, height=6cm]{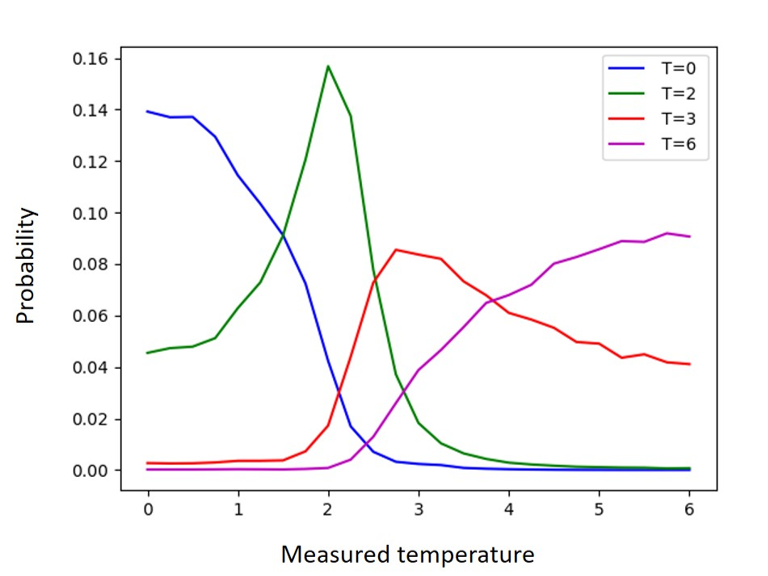}
\caption{\small Probability distributions of measured temperatures for various sets of configurations 
generated at $T=0,2,3,6$ respectively. Temperature of the configurations can be distinguished
by looking at the shapes of the distributions.}
\label{fig_temp}
\end{center}
\end{figure}

In Fig.\,\ref{fig1}, we plot behaviors of the loss function (\ref{loss_sl})
as we iterate renewals of the weights and biases (\ref{WBrenewal-TM}).
The blue {(lower)} line  shows the 
training error, namely
values of the loss function (\ref{loss_sl}) after 
iterations of training 
using  25000 configurations.
It is continuously decreasing, even after 7500 epochs.
On the other hand, the red {(upper)} line shows the test error, namely
values of the loss function for additional 25000 configurations which are not used for the training.
This is also decreasing at first, but after 6000 epochs it becomes almost constant.
After 7500 epochs, in fact, it turns to increase.
This means the machine becomes over-trained, 
therefore we stopped the learning at 7500 epochs. 

In Fig.\,\ref{fig_temp} we show  probability distributions of temperature this NN measures.
Here we use configurations at $T=0,2,3,6$ which are not used for the training.
Though they are not sharply peaked at the temperatures where the configurations are generated\footnote{
\label{footnote:broad}
There are two reasons for this broadening of the distributions.
One is due to the finiteness of the size of a configuration $N=L \times L=10 \times 10.$ 
Another is due to the limit of measuring temperature by the NN. 
If the size of a configuration was infinite and if the ability of discriminating subtle differences
of different temperature configurations was limitless, we would have obtained a very sharp peak
at the labelled temperature.}, 
each of them has characteristic shape that is different temperature by temperature.
{Thus it is possible to distinguish the temperature of the input configurations 
by looking at the shape of the probability distribution, even if these configurations 
are not used for the training of the NN.}

In the following, by using this NN, 
we measure temperature of configurations that
are generated by the RBM flow.  
 

\subsection{Unsupervised RBM flows}
\label{sec:res_rbm}

Now we present the main results of the present paper, namely 
 the flows generated by the unsupervised RBM.  {We sometimes call it the RBM flow.}
As discussed in the Introduction, 
if the RBM is similar to the the conventional RG in that it possesses 
 a  function of coarse-graining,
the RBM flow must go  away from the critical point  $T_c=2.27$. 
In order to check it
we construct three different types of unsupervised RBMs, which we call 
type V, type L, and type  H respectively, 
using the method of Sec.\,\ref{method-USRBM}.
Each of them is trained by a different set of 
spin configurations generated at different set of temperatures. 
We then generate flows of temperature distributions  by using  these trained RBMs,  
following the methods of Secs.\,\ref{method-Flow} and \ref{sec:NN_set}.

\subsubsection*{Type V RBM: Trained by configurations at $T=\{0,\ 0.25,0.5, \cdots, 6\} $}

First we construct type V RBM, which is trained 
by configurations at temperatures ranging widely from low to high, $T=0,0.25,\ldots,6$.
The temperature range includes the temperature $T=2.25 $ near $T_c$.
After training is completed,
this unsupervised RBM will have learned  features of spin configurations at these
temperatures.

Once the training is finished, 
we then generate a sequence of reconstructed configurations as in Eq.\,(\ref{config-flow-v})
using the methods in Sec.\,\ref{method-Flow}.
For this, we prepare two different sets of initial configurations. 
One is a set of configurations 
at $T=0$, and another at $T=6$. These initial configurations are not used for the training of the RBM. 
Then by using the supervised NN in Sec.\,\ref{sec:res_sl}, we measure  temperature and
translate the flow of configurations to a flow of temperature distributions.

\begin{figure}[h]
\begin{center}
\includegraphics[width=8cm, height=6cm]{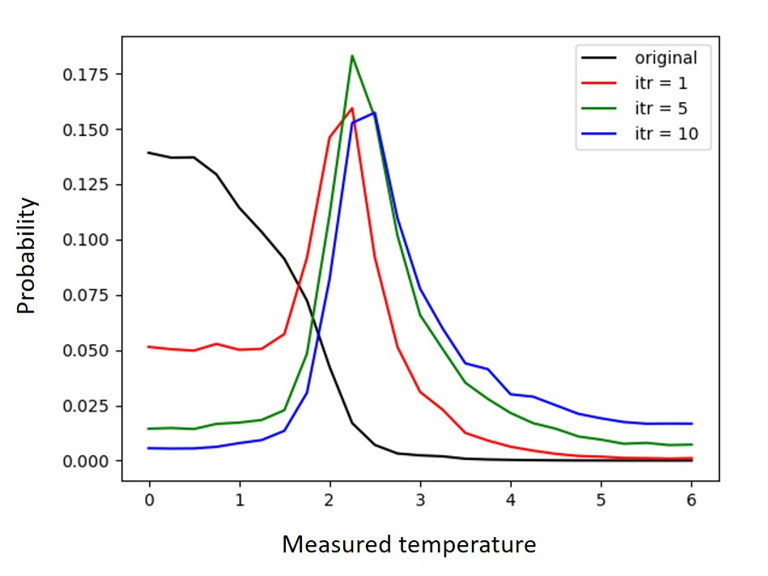}
\includegraphics[width=8cm, height=6cm]{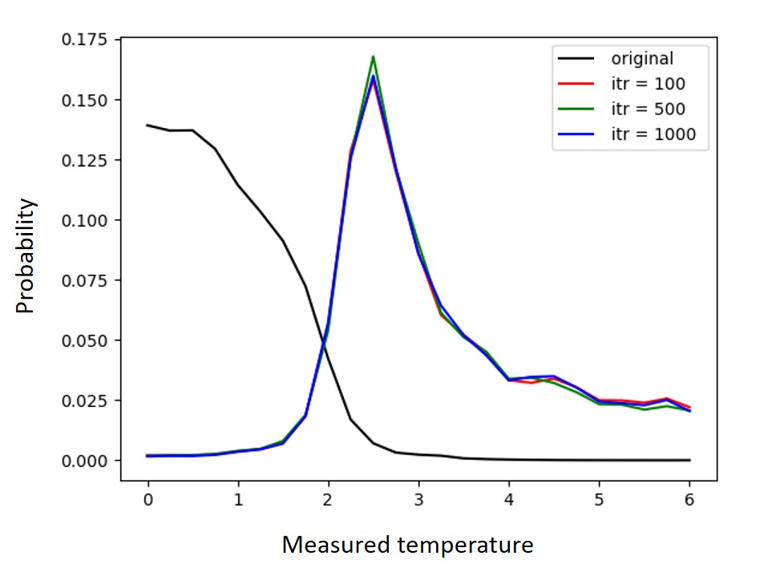}
\caption{\small 
Temperature distributions after various numbers of iterations of type V RBM,
which is trained by the configurations at $T=0,0.25,\ldots,6$.
The original configurations are generated at $T=0$. 
{After only several iterations, the temperature distribution is peaked around $T_c$,
and stabilize there: $T_c$ is a stable fixed point of the flow.}}
\label{fig2}
\end{center}
\end{figure}

\begin{figure}[!h]
\begin{center}
\includegraphics[width=8cm, height=6cm]{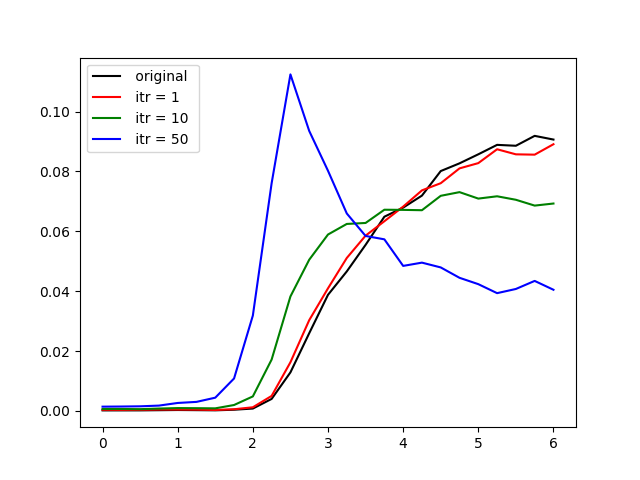}
\includegraphics[width=8cm, height=6cm]{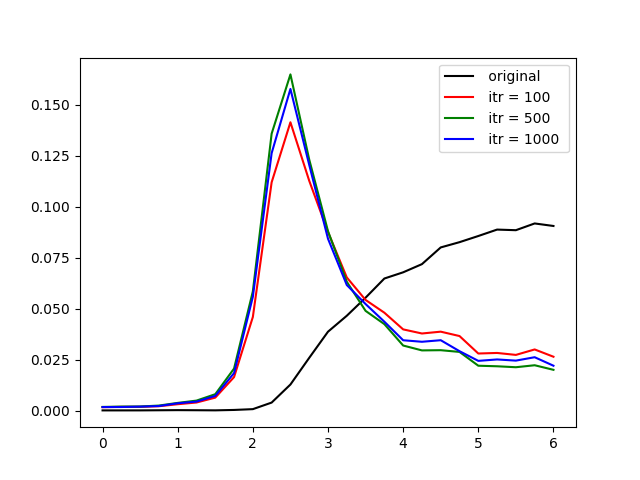}
\caption{\small 
Temperature distributions after various numbers of iterations of 
{the same RBM as Fig.\,\ref{fig2}.}
The original configurations are generated at $T=6$. 
{After $\sim 50$ iterations, the distribution
stabilizes at $T_c$.}}
\label{fig2_6}
\end{center}
\end{figure}

In Figs.\,\ref{fig2} and \ref{fig2_6}, we plot temperature distributions of configurations
that are generated by  iterating the RBM reconstruction  
in Sec.\,\ref{method-Flow}. 
The ``itr" in the legends means the numbers of iterations $n$ by the unsupervised RBM. 
Fig.\,\ref{fig2} 
shows a flow of temperature distributions starting from spin configurations generated at $T=0$. 
Fig.\,\ref{fig2_6} 
starts from $T=6.$ 
In all the figures, the black lines are the measured temperature distributions of the 
{initial} configurations\footnote{
As discussed in the footnote \ref{footnote:broad}, these distributions are not sharply peaked at the
temperature at which the configurations are generated.}.
Colored lines show temperature distributions of 
the reconstructed configurations  $\{v_i^{(n)} \}$ after various numbers of iterations.
{The left panels show the temperature distributions at small iterations
 (up to 10 in Fig.\,\ref{fig2} and 50 in Fig.\,\ref{fig2_6}), while the
right panels are at larger iterations up to 1000.   
These results indicate that the critical temperature $T_c$
is a stable fixed point of the flows in type V RBM.}
{It is apparently  different from a naive expectation 
that the RBM flow should show the same behavior as the RG flow.}
Indeed it is in the opposite direction. 
From whichever temperature $T=0$ or $T=6$ we start the RBM iteration,
the peak of the temperature distributions 
approaches the critical point ($T=2.27$). 

\begin{figure}[h]
\begin{center}
\includegraphics[width=8cm, height=6cm]{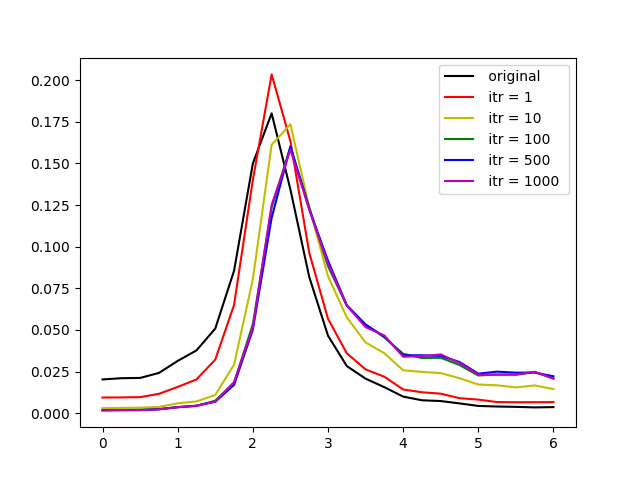}
\caption{\small 
Temperature distributions after various numbers of iterations of the same RBM as Figs.\,\ref{fig2} and \ref{fig2_6}. 
The original configurations are generated at $T=2.25$. 
The distribution is stable at around $T_c$.}
\label{fig3}
\end{center}
\end{figure}

In order to confirm the above behavior, we provide another set of configurations 
at $T=2.25$ as initial configurations, and generate the flow of temperature
by the same trained RBM.
The flow of temperature distributions is shown in Fig.\,\ref{fig3}. 
We can see that 
the temperature distribution of the reconstructed configurations 
remains near the critical point, and never flows away from there\footnote{We also trained the RBM using configurations of wider range 
of temperatures; $T=0,0.25,\ldots,10,\infty$.
The results are very similar, and the temperature distributions 
of reconstructed configurations always approach the critical point $T_c=2.27$.}. 
If the process of the unsupervised RBM corresponds to
coarse-graining of spin configurations, the temperature distributions of the
reconstructed configurations must flow away from $T_c$.
Though the direction of the flow is opposite to the RG flow, 
both flows have the same {property 
in that the critical point $T=T_c$ plays an important role in controlling the flows.}

So far,  in obtaining the above results of Figs.\,\ref{fig2}, \ref{fig2_6} and \ref{fig3},
we used an unsupervised RBM with 64 neurons in the hidden layer. 
We also trained other RBMs with different sizes of the hidden layer,
but by the same set of spin configurations.
When the size of the hidden layer is smaller than {(or equal to)}
that of the visible layer $N_v=100$, namely 
$N_h = 100, 81, 64, 36$ or $16$, we find that 
the temperature distribution  approaches the critical point.
A difference is that 
{for smaller $N_h$,}
the speed of the flow to approach $T_c$ becomes faster 
(i.e.,  the flow arrives at $T_c$ by smaller numbers of iterations). 

\begin{figure}[!h]
\begin{center}
\includegraphics[width=8cm, height=6cm]{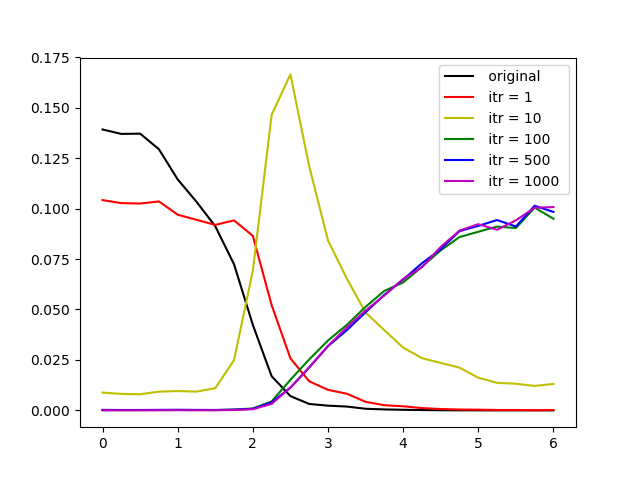}
\caption{\small 
Temperature distribution after various numbers of iterations of type V RBM with 225 neurons in the hidden layer; i.e., $N_h>N_v$. The original configurations are generated at $T=0$.
The distribution has a peak at $T=T_c$ after 10 iterations, but then moves towards $T=\infty$.}
\label{fig4}
\end{center}
\end{figure}

In contrast, 
when the RBM has more than 100 neurons in the hidden layer; $N_h> N_v$,
 we  obtain different results.  
Fig.\,\ref{fig4} shows the case of $N_h=225$ neurons. 
Until about ten iterations, the measured temperature distribution
 behaves similarly to the case of $N_h\leq 100$, i.e.,
 it approaches the critical temperature.
However, afterward  it passes the critical point and flows away to higher temperature. 
In the case of 400 neurons, it moves towards high temperature at faster speed. 
This behavior suggests that, if the hidden layer has  more than a necessary size, the NN tends to
learn {a lot of noisy fluctuations.}
Since  configurations at higher temperatures are noise-like,
the flow should go away to high temperature.
We come back to this conjecture in later sections.

\subsubsection*{Type H/L RBM: Trained by configs at Higher/Lower temperatures}
\begin{figure}[!h]
\begin{center}
\includegraphics[width=8cm, height=6cm]{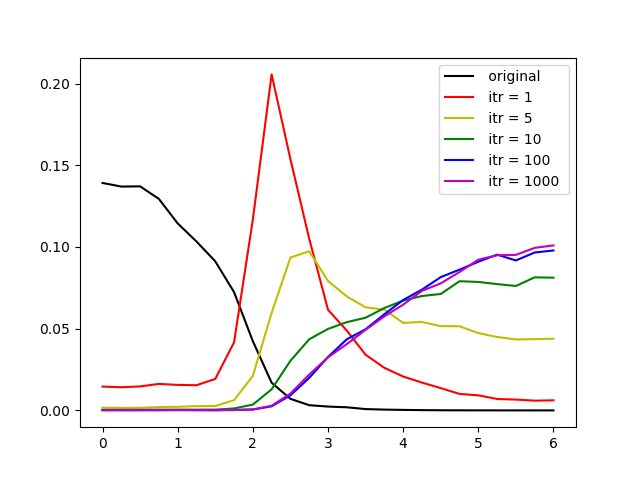}
\includegraphics[width=8cm, height=6cm]{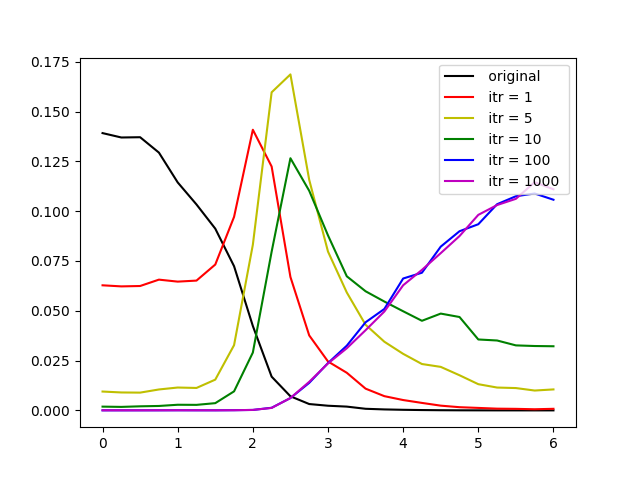}
\caption{\small Flow of temperature distributions starting from $T=0$ in type H RBM.
Type H RBM is trained by configurations at only $T=4, 4.25,\ldots,6$. 
The NN has $N_h=64$ neurons (left) and $N_h=225$ neurons (right) respectively in the hidden layer.
The speed of the flow is slower for the larger sized hidden layer.}
\label{fig5}
\end{center}
\end{figure}
\begin{figure}[!h]
\begin{center}
\includegraphics[width=8cm, height=6cm]{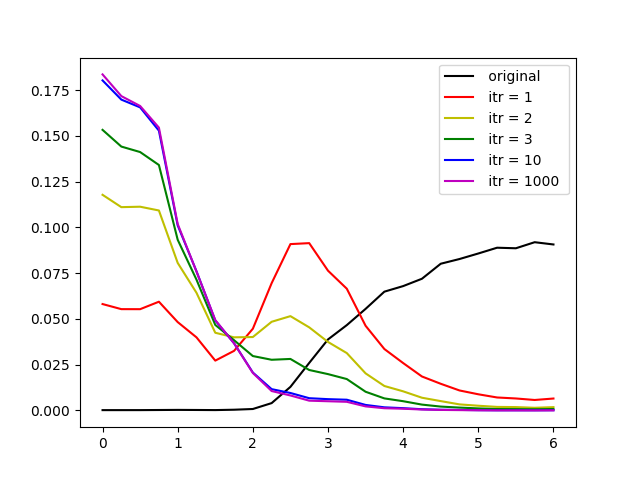}
\includegraphics[width=8cm, height=6cm]{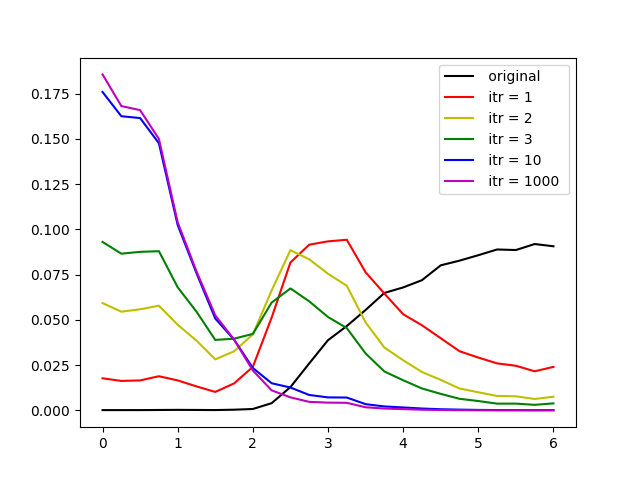}
\caption{\small Flow of temperature distributions starting from $T=6$ in type L RBM.
Type L RBM is trained by configurations at only $T=0$.  
$N_h=64$  (left) and $N_h=225$  (right).
}
\label{fig6}
\end{center}
\end{figure}

Next we construct another type of RBM, which is trained by configurations at higher temperatures $T=4, 4.25, \ldots, 6$
than $T_c \sim 2.25.$ We call it type H RBM. 
The results of the flows of temperature distributions in type H RBM
are drawn in Fig.\,\ref{fig5}. 
In this case, 
the measured temperature passes the critical point
and goes away towards higher temperature.
{The behavior is understandable since the RBM must have learned
only the features at higher temperatures.}
We also find that, if the number of neurons in the hidden layer is increased,
the  {flow moves}  more slowly.

Finally, we construct  type L RBM, which is trained by configurations
only at the lowest temperature $T=0$. 
Fig.\,\ref{fig6} shows the numerical results of  flows in the type L RBM.
Similarly to the type H RBM, 
the measured temperature passes the critical point,
but flows towards lower temperature instead of higher temperature.
It is, of course, as expected because the type L RBM must have learned the features 
of spin configurations at $T=0$. 
In the type L RBM, as far as we have studied, 
the flow never goes back to higher temperature even for large $N_h$. 
It will be because the $T=0$ configurations used for training 
 do not at all contain noisy fluctuations specific to high temperatures.
{This also suggests that the RBM does not learn features that are not contained 
in the configurations used for trainings.}

\subsubsection*{Summaries and Conjectures}

Here we first summarize the numerical results:

\vspace{3mm}

\noindent
For the type V RBM,
\begin{itemize}
\item 
When $N_h \leq 100=N_v$, 
the measured temperature $T$ approaches $T_c$
(Figs.\,\ref{fig2}, \ref{fig2_6} and \ref{fig3}).
\item However, for $N_h > 100=N_v$, the flow eventually goes away towards $T=\infty$
(Fig.\,\ref{fig4}).
\item Speed of  flow 
is slower for a larger $N_h$.
\end{itemize}
For the type H/L RBM,
\begin{itemize}
\item The temperature $T$ flows towards $T=\infty / T=0$ respectively
(Figs.\,\ref{fig5} and \ref{fig6}).
\item Speed of  flow 
 is slower for a larger $N_h$.
\end{itemize}
Here $N_h$ and $N_v$ are numbers of hidden and visible neurons {in the RBM}.
These behaviors are  reflections of the properties of the weights and biases that
the unsupervised RBMs have learned in the process of training. 

\vspace{3mm}

{Understanding the above behaviors is equivalent to answering
what  the unsupervised RBMs have learned in the process of trainings.
The most important question will be  
why the temperature approaches $T_c$  in the type V RBM with $N_h \le N_v$,
instead of, e.g., broadening over the whole regions of temperature from $T=0$ to $T=6$.}
Note that we did not teach the NN neither about the critical temperature nor the presence of phase transition.
We just have trained the NN by configurations at various temperatures, from $T=0$ to $T=6$.
Nevertheless the numerical simulations show that 
 the temperature distributions are peaked at $T_c$ after some iterations of the RBM reconstruction.
Thus we are forced to conclude that 
the RBM has automatically learned  {\it features} specific to the critical temperature $T_c$.

An important feature at $T_c$ is the scale invariance.
We have generated spin configurations at various temperatures by the Monte Carlo method,
and each configuration has typical fluctuations specific to each temperature.
At very high temperature, fluctuations are almost random at each lattice site
 and there are no correlations between spins at {distant} positions. 
At lower temperature, they become correlated: 
 the correlation length becomes larger  as $T \rightarrow T_c$ and diverges at $T_c$.
On the other hand, at $T \ll T_c$, spins are clustered and in each domain all spins take 
$\sigma_{x,y}=+1$ or $\sigma_{x,y}=-1$. 
At low temperature configurations have only big clusters, and as 
 temperature increases small-sized clusters appear.
At $T_c$, spin configurations become to have clusters of various sizes in a scale-invariant way. 

Now let us come back to the question
why the type V RBM generates a flow  approaching $T_c$ {and does not 
randomize to broaden the temperature distribution over the whole regions.}
We have trained the type V RBM by using configurations at various temperatures with 
different {sized clusters}, 
and in the process the machine must have simultaneously acquired features at various temperatures.
Consequently the process of the RBM reconstruction adds various features that the machine
has learned to a reconstructed configuration.
If  only  a single feature at a specific temperature  was added to the reconstructed configuration, 
the distribution  would   become to have a peak at
this temperature. 
But it cannot happen because various features of different temperatures will be
added to a single configuration by iterations of reconstruction processes.  
Then one may ask if 
 there is a configuration that is  stable
under additions of features at various different $T$.

Our first conjecture about this question is that
a set of configurations at $T_c$ is a stabilizer (and even more an attractor)
of the type V RBM with $N_h\leq N_v$. 
It {must be} due to the scale invariant properties of the configurations at $T_c$.
Namely since these configurations  are scale invariant,
they  have  all the {\it features} of  various temperatures simultaneously, and consequently
they can be the  stabilizer of   this  RBM. 
This sounds plausible since  the scale invariance means that the configurations have
various different characteristic length scales.
However, we notice that
 this doesn't mean that the RBM has forgotten the features of configurations away from the critical point.
Rather it means that the RBM has learned features of all temperatures simultaneously.
This doesn't mean either that the configurations at $T=T_c$ have especially affected 
strong influence on the machine in the process of training. 
It can be confirmed as follows.
Suppose we have trained a RBM by configurations at temperatures excluding $T=T_c$,
namely train by configurations at all  temperatures {\em except} $T=2.25$ and $2.5$.
We found in the numerical simulations 
that the RBM generates a flow towards the critical point though we did not
provide configurations at $T=T_c$. 
Therefore we can say that the type V RBM has learned the features at all the temperatures
and that configurations at $T_c$ are special  because they contain
all the features of various temperatures  in the configurations.

{Our second conjecture, which is related to the behavior of  the type V RBM with $N_h > N_v$, 
is  that RBMs with unnecessary large sized hidden layer tend to
learn lots of irrelevant features.
In the present case,  they are noisy fluctuations of configurations at high temperatures.}
High temperature configurations have only short distance correlations, whose behavior is
similar to the typical behavior of noise.  
The conjecture will be partially supported by 
the similarity of the RBM flows between the
type V RBM with $N_h > N_v$ and the type H RBM. 
Namely both RBM flows  converge on $T=\infty$.
The similarity indicates that 
the NN with a larger number of $N_h$ may have learned too much
noise-like features of configurations at higher temperatures.
The above considerations will teach us that 
the moderate size of the hidden layer, $N_h < N_v$, is the most efficient 
to properly extract the features.

\section{Analysis of the weight matrix}
\label{weights}

In the previous section, we showed our numerical results for the flows generated by 
unsupervised RBMs,  
and proposed two conjectures.
One is that the {scale invariant} $T=T_c$ configurations are  {stabilizers}  
of the type V RBM flow.
{Another conjecture is  that the RBM with an unnecessary large sized hidden layer $N_h>N_v$ 
tends to learn too much irrelevant noises.}
In this section, to further understand the theoretical basis of feature extractions
and to give supporting evidences for our conjectures,  
{we analyze various properties of the weight matrices and biases of the trained RBMs. 
Particularly, we study properties of $WW^T$ by looking at spin correlations 
in Sec.\,\ref{sec:correlation}, magnetization in Sec.\,\ref{sec:SVD}, and eigenvalue spectrum in Sec.\,\ref{sec:spectrum}.}


\subsection{Why $WW^T$ is important}
\label{sec:WWT}
All the information  that the machine has learned 
is contained in the weights $W_{ia}$ and the biases $b_i^{(v)}, b_a^{(h)}$.
Since the biases have typically  smaller values than the weights (at least in the present situations), 
we will concentrate on the weight matrix $W_{ia}$ ($i=1, \ldots, N_v=L^2$; $a=1, \ldots, N_h$) in the following.

Let us first note that the weight matrix $W_{ia}$ transforms as
\ba
W_{ia} ~\rightarrow~ \tilde{W}_{ia} =\sum_{j,b} U_{ij} W_{ib} (V^T)_{ba}
\ea
under transformations\footnote{Since spin variables on each lattice site are restricted to take 
values $\pm 1$, the matrices, $U_{ij}$ and $V_{ab}$, are elements of the
symmetric group, not the orthogonal Lie group.} of exchanging 
the basis of neurons in the visible layer ($U_{ij}$) and in the hidden layer ($V_{ab}$).
Since the choice of basis in the hidden layer is arbitrary, relevant information in the visible layer
is stored in a combination of $W_{ia}$ that is invariant under  transformations of $V_{ab}$.
The simplest combination is a product,
\ba \label{WW}
(WW^T)_{ij} = \sum_a W_{ia} W_{aj}\,.
\ea
It is an $N_v \times N_v = 100\times 100$ matrix, and  independent of the size of $N_h$.
But its property depends on $N_h$ because
the rank of $WW^T$ must be always smaller than $\min (N_v, N_h)$.
Thus, if $N_h<N_v$, the weight matrix is strongly constrained; e.g. a unit matrix $WW^{T}=1$ is not allowed. 

This simplest product 
 (\ref{WW})  plays an  important role in the dynamics of the flow generated by the RBM.
 It can be shown as follows. 
If the biases are ignored, the conditional probability (\ref{p_model}) 
and the expectation value (\ref{vev-h-inv}) for $h_a$ in the background of $v_i$ become
\ba
p(\{h_a\}|\{v_i\}) = \frac{e^{\sum_i v_i W_{ia} h_a}}{2\cosh \left( \sum_i v_i W_{ia}\right) }
\,,\qquad
\langle h_a \rangle = \tanh \left(\sum_i v_i W_{ia}\right)\,.
\ea
In $p(\{h_a\}|\{v_i\})$, 
a combination $\sum_i v_i W_{ia} =: B_a$ can be regarded as an external magnetic field for $h_a$.
Thus these two variables, $B_a$ and $h_a$,  tend to correlate with each other.
Namely,  the probability $p(\{h_a\}|\{v_i \})$ becomes larger when they have the same sign. 
Moreover, for $|B_a|<1$, $\langle h_a \rangle$ is approximated by $B_a$  and
we can roughly identify these two variables,
\ba
\langle h_a \rangle \sim B_a := \sum_i v_i W_{ia}\,.
\label{hBalign}
\ea
It is usually not a good approximation since weights can have larger values, 
{but let us assume this for the moment.}
For a large value of $|B_a|$, $\langle h_a \rangle$ is saturated at $\langle h_a \rangle= B_a/|B_a|$.

Suppose that the input configuration is given by $\{v_i^{(0)} \} = \{\sigma_i^{A} \}$.
If Eq.\,(\ref{hBalign}) is employed, we have
$h_a^{(1)} = B_a^{(0)} = \sum_i v_i^{(0)} W_{ia} .$
Then the conditional probability (\ref{p_tilde}) in the background of $ h_a^{(1)} $ with $b_a^{(h)}=0$, 
\ba
 p(\{ v_i\}|\{h_a^{(1)}\}) = \prod_i \frac{e^{\sum_{a}  v_i W_{ia} h_a^{(1)}}}{2\cosh \left( \sum_a W_{ia} h_a^{(1)} \right)} 
\ea
can be approximated as
\ba
 p(\{ v_i\}|\{ h_a^{(1)} \})
\sim \prod_i
\frac{e^{ \sum_{a}  v_i W_{ia} \sum_j W_{ja} v_j^{(0)}}}{2\cosh \left( \sum_a W_{ia} \sum_j W_{ja} v_j^{(0)} \right) }
= \prod_i \frac{e^{\sum_j  v_i (WW^T)_{ij} v_j^{(0)} }}{2 \cosh \left( \sum_{j} (WW^T)_{ij} v_j^{(0) }\right) }\,.
\label{approx-p(v)}
\ea
The RBM learns the input data $\{v_j^{(0)}\}$  
so that the probability distribution  $p$ reproduces the probability distribution
of the initial data, $q(\{v_i\})=\frac{1}{N} \sum_A \delta(v_i - \sigma_i^{A})$.
{Therefore, training of the RBM will be performed  so as to
 enhance  the value  $\sum_{A,i,j} \sigma_i^{A(0)} (WW^T)_{ij} \sigma_j^{A(0)}$. 
 This means that $W$ is chosen so that $(WW^T)_{ij}$  will reflect the spin correlations
  of the input configurations $\{ \sigma_i^{A} \}$ at site $i$ and $j$.}

In this simplified discussion, 
learning of the RBM is performed through  the combination $WW^T$. 
Of course, we neglected 
 the nonlinear property of the neural network 
  and the above statement cannot be justified as it is. 
  Nevertheless, we will find below that the analysis of $WW^T$ is quite useful to understand 
how the RBM works. 


\subsection{Spin correlations in $WW^T$}
\label{sec:correlation}
\begin{figure}[h]
\begin{center}
\includegraphics[width=5cm, height=4.6cm]{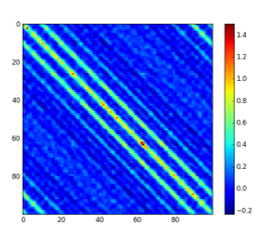}
\includegraphics[width=5cm, height=4.6cm]{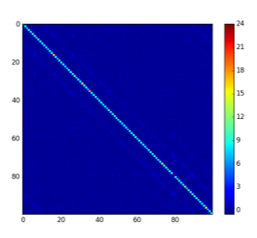}
\includegraphics[width=5cm, height=4.6cm]{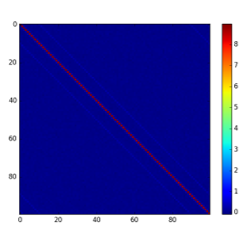}
\caption{\small Elements of $WW^T$ when the hidden layer has 16 (left), 100 (center), 400 (right) neurons.}
\label{fig7}
\end{center}
\end{figure}

In Fig.\,\ref{fig7}, we plot values of 
matrix elements of the $100\times 100$ matrix $WW^T$. 
These three figures correspond to the RBMs with different sizes of $N_h$.
We can see that they have large values in the diagonal and near diagonal elements.
Note that the spin variables in the visible layer, $\sigma_{x,y}$ with $x,y=1, \ldots, L=10$, 
are lined up as 
$(\sigma_{1,1}, \sigma_{1,2}, \ldots, \sigma_{1,L}, \sigma_{2,1}, \ldots, \sigma_{2,L}, \sigma_{3,1}, \ldots, \sigma_{L,L})$, 
and  named $(\sigma_1, \sigma_2, \ldots, \sigma_N)$. 
Hence  lattice points $i$ and $j$ of   
$\sigma_i$ ($i=1,\ldots, L^2$) are adjacent to each other when $j=i \pm 1$ and $j=i \pm L$. 
In the following, we mostly discuss the type V RBM unless otherwise stated.

As discussed above, the product of weight matrices $WW^T$ must reflect correlations
between spin variables of the input configurations  used for the training of the RBM.
The most strong correlation in $v_i^{(0)} (WW^T)_{ij} v_j^{(0)}$ 
 is of course the diagonal component, $i=j$. 
Thus we expect that the matrix
 $WW^T$ will have large diagonal components.
Indeed, {such} 
behavior can be seen in Fig.\,\ref{fig7}. 
In particular, for $N_h=400 > N_v=100$ (the rightmost figure), 
$WW^T$ is clearly close to a diagonal matrix.
It is  almost true for the case of $N_h=100=N_v$ (the middle figure).
However, for $N_h=16 < N_v=100$ (the leftmost figure), it is different from a unit matrix
and off-diagonal components
of $(WW^T)_{ij}$ also have large values, in particular, 
 at $j=i+1$ and $j=i+2$. 
 This behavior must be a reflection
 of the spin correlations of the input configurations \footnote{Off-diagonal components 
of $j=i+L$ or $j=i+2L$ are also large, which corresponds to  correlations along $y$-direction.
Large off-diagonal components at $j=i+1$ and $j=i+2$ mean correlations along $x$-direction.}. 
It is  also a reflection of the fact that 
 the rank of $WW^T$ is smaller than $N_h$ and 
  $WW^T$ cannot be a unit matrix if $N_h < N_v$.
 Thus even though only less information 
 can be stored in the weight matrix for a smaller number of hidden neurons,
the relevant information of the spin correlations is well encoded 
in the weight matrix of the RBM with $N_h<N_v$ compared with the RBM with larger $N_h$.
Then we wonder why such relevant information is  lost in the RBM with $N_h > N_v$.
This question might be related to our second conjecture  proposed at the end of 
Sec.\,\ref{sec:res_rbm} that the RBM with very large $N_h$ will learn 
too much  irrelevant information, namely noises of the input configurations. 
It is interesting and a bit surprising that
the RBM with fewer hidden neurons seems to learn
more efficiently the relevant information of the spin correlations. 

\begin{figure}[h]
\begin{center}
\includegraphics[width=8cm, height=6cm]{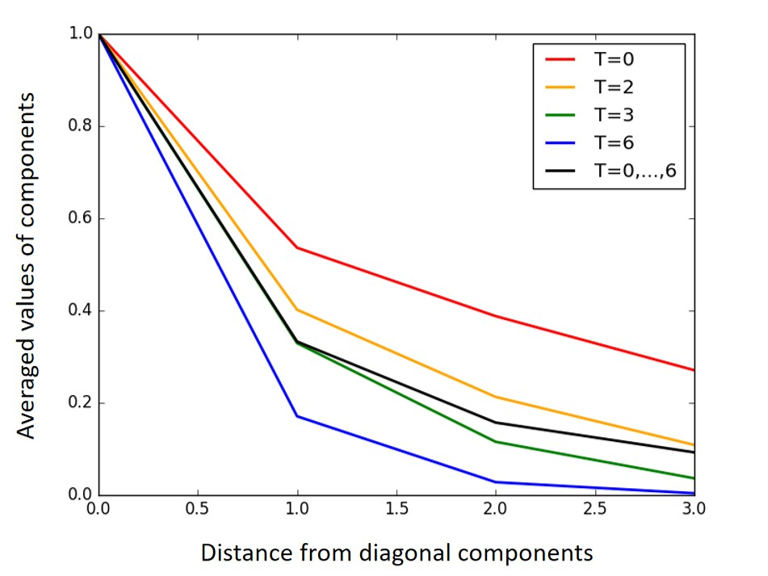}
\caption{\small Averaged values of the off-diagonal components of $WW^T$ (normalized by the diagonal components). 
Each colored line corresponds to the RBM that has learned
 configurations at a single temperature $T=0, 2, 3, 6$ respectively.
The black line  (the most middle line)  
 is the behavior of the type V RBM that has learned all the temperatures $T=0,\ldots,6$.}
\label{fig8}
\end{center}
\end{figure}

In order to further confirm the relation between 
 the correlations in the combination of the weight matrix $WW^T$  and the spin correlations
 of the input configurations, 
 we will study  structures of the weight matrices of 
 other types of RBMs.  
In Fig.\,\ref{fig8}, we plot  behaviors of the off-diagonal components of  $WW^T$
for various RBMs.
Each RBM is trained by configurations at a single temperature $T=0$ (type L), $T=2$, $T=3$ and $T=6$ respectively.
The size of the hidden layer is set to $N_h=16$.
For comparison, we also plot the behavior of the off-diagonal components for the type V RBM.

Fig.\,\ref{fig8} shows that the correlation of $WW^T$ decays more rapidly 
at higher temperature, which is consistent with the expected behavior of spin correlations.
Therefore, the RBM seems to learn correctly about the correlation length, or the size of clusters, which becomes smaller at higher temperature.
Furthermore, we find that,
for the type V RBM that has learned all temperatures $T=0,\ldots,6$,
the off-diagonal elements decrease with the decay rate between the $T=2$ case and the $T=3$ case. 
This indicates  that 
{the type V RBM  has acquired similar features to those of the configurations around $T_c=2.27$.
It is consistent with the numerical results of  Figs.\,\ref{fig2}, \ref{fig2_6} and \ref{fig3},
and gives another circumstantial evidence supporting for the first conjecture in Sec.\,\ref{sec:res_rbm}.}

\subsection{Magnetization and singular value decomposition (SVD)}
\label{sec:SVD}

Information of the weight matrix $W$ can be inferred by using the method of the
 singular value decomposition (See, e.g., \cite{Lee:2014ama, 1608.08333}).
Suppose that the matrix $WW^T$ has eigenvalues $\lambda_a$ ($a=1, \ldots, N_v)$
 with corresponding eigenvectors $u_a$;
\ba
WW^T u_a = \lambda_a u_a\,.
\ea
Decomposing an input configuration vector $v^{(0)}$ in terms of the eigenvectors $u_a$ as
$v^{(0)} = \sum_a c_a u_a$ with a normalization condition $\sum_a (c_a)^2=1$, 
we can rewrite $v^{(0)T} WW^T v^{(0)}$ as 
\ba
v^{(0)T} WW^T v^{(0)} = \sum_a c_a^{2} \lambda_a\,.
\ea
Thus if a vector $v^{(0)}$ contains more components with larger eigenvalues of $WW^T$,
the quantity $v^{(0)T} WW^T v^{(0)}$ becomes larger. 

\begin{figure}[h]
\begin{center}
\includegraphics[width=8cm, height=6cm]{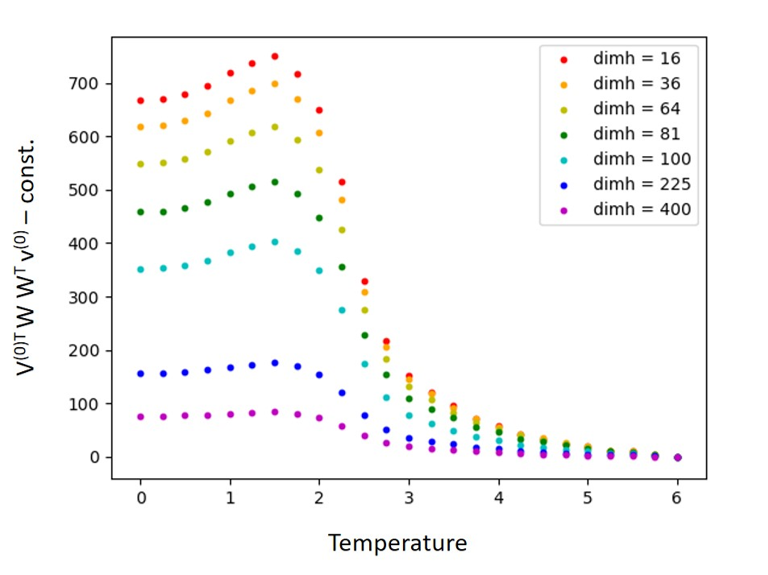}
\caption{\small
Averaged values of $v^{(0)T} WW^T v^{(0)}$ over the 1000 input configurations at each temperature.
Different colors correspond to type V RBMs with different number of hidden neurons $N_h$.
In this figure, the values at $T=6$ are subtracted for comparison between different RBMs.}
\label{fig9V}
\end{center}
\end{figure}

Fig.\,\ref{fig9V} shows averaged values of $v^{(0)T} WW^T v^{(0)}$
over the 1000 configurations $\{v^{(0)} \}$ at each temperature. 
For comparison between different RBMs, we subtracted 
the values at $T=6$.  
The figure  shows a big change near the critical point, which is reminiscent of 
 the magnetization of Ising model.
{Since $v^{(0)T} WW^T v^{(0)}$  should contain more
 information than the magnetization itself, the behavior cannot be exactly the same.
 But it is quite  intriguing that Fig.\,\ref{fig9V} shows  similar behavior to the magnetization\footnote{
 The behavior indicates that the principal eigenvectors
with large eigenvalues might be related to the magnetization, and 
information about the phase transition is surely imported in the weight matrix. 
Thus we  investigated properties of the eigenvectors but so far we 
have not got  any physically reasonable pictures.
 We want to come back to this problem in future works.}.
It might be because the quantity contains much information about the lower temperature
after subtraction of the values at higher temperature\footnote{
It suggests that the subtraction may correspond to removing the contributions 
of the specific features at higher temperature.}.}

\begin{figure}[h]
\begin{center}
\includegraphics[width=8cm, height=6cm]{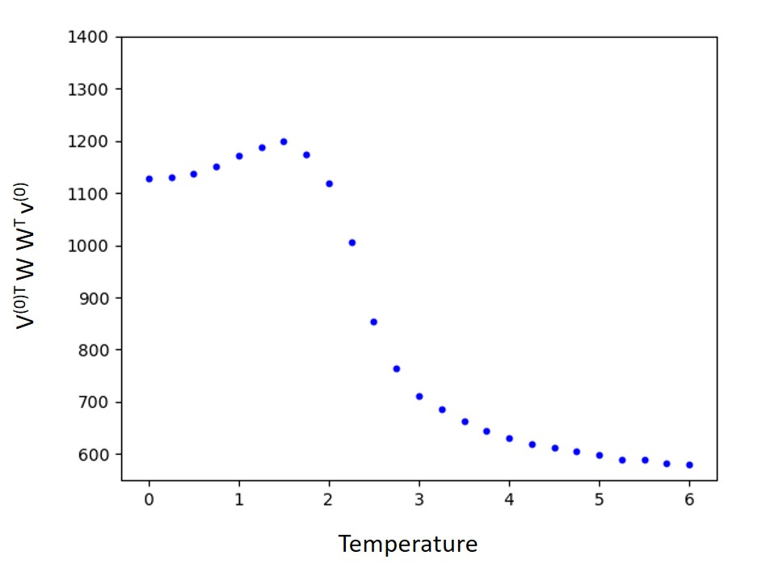}
\includegraphics[width=8cm, height=6cm]{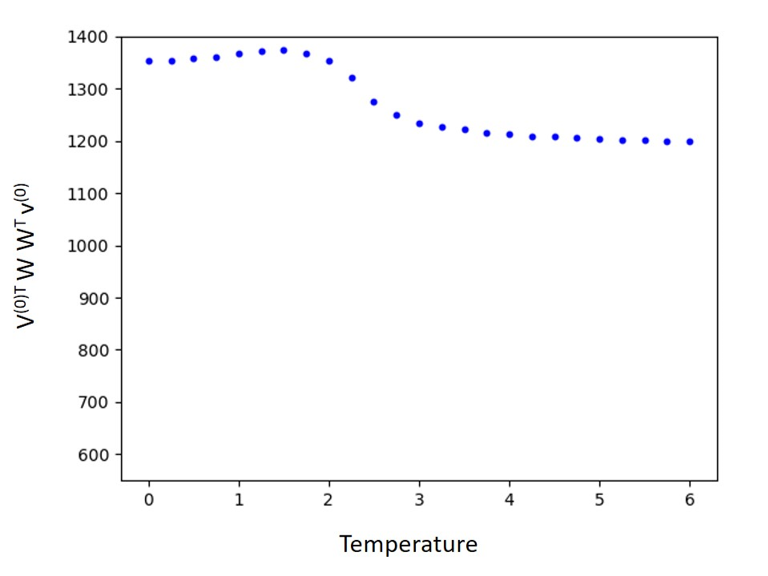}
\caption{\small 
Averaged values of $v^{(0)T} WW^T v^{(0)}$ over the 1000 configurations $\{v^{(0)} \}$ at each temperature. 
The left and right figure shows the quantities in the type V RBM 
 with $N_h=64$ and $N_h=225$ respectively.}
\label{fig9}
\end{center}
\end{figure}

In order to see the properties of $v^{(0)T} WW^T v^{(0)}$ more than the magnetization in Fig.\,\ref{fig9V},
we plot the same quantities but without subtracting the values at $T=6$. 
Fig.\,\ref{fig9} shows two cases for $N_h=64$ and $N_h=225.$
{These figures show
that, at high temperature,
the RBM with large $N_h$  in the right panel has  
 larger components of the principal eigenvectors} 
 compared to the RBM with small $N_h$ in the left panel.
The difference must have caused the different behaviors 
in the RBM flows  shown in Fig.\,\ref{fig2} ($N_h=64)$ and Fig.\,\ref{fig4} ($N_h=225$).
Namely the former RBM flow approaches the critical temperature $T_c$,
while the latter eventually goes towards  higher temperature.
 The difference of two figures in Fig.\,\ref{fig9V} indicate that 
the  RBM with larger $N_h$ seems to have learned more characteristic
 features at high temperatures than  the RBM with fewer $N_h$.
Then, does the RBM with small $N_h$ fail to  learn the features of high temperatures? 
Which RBM is more adequate for feature extractions? 
Although it is difficult to answer which is more adequate without specifying
what we want the machine to learn, 
we believe that  the RBM with $N_h < N_v$ 
properly learns all the features of various temperatures 
while  the RBM with $N_h>N_v$ has learned too much irrelevant
features of high temperature.
This is nothing but the second conjecture in Sec.\,\ref{sec:res_rbm}, and
supported by the behaviors of correlations in $WW^T$ 
discussed in Sec.\,\ref{sec:correlation}.

\subsection{Eigenvalue spectrum and information stored in $W$}
\label{sec:spectrum}

\begin{figure}[h]
\begin{center}
\includegraphics[width=8cm, height=6cm]{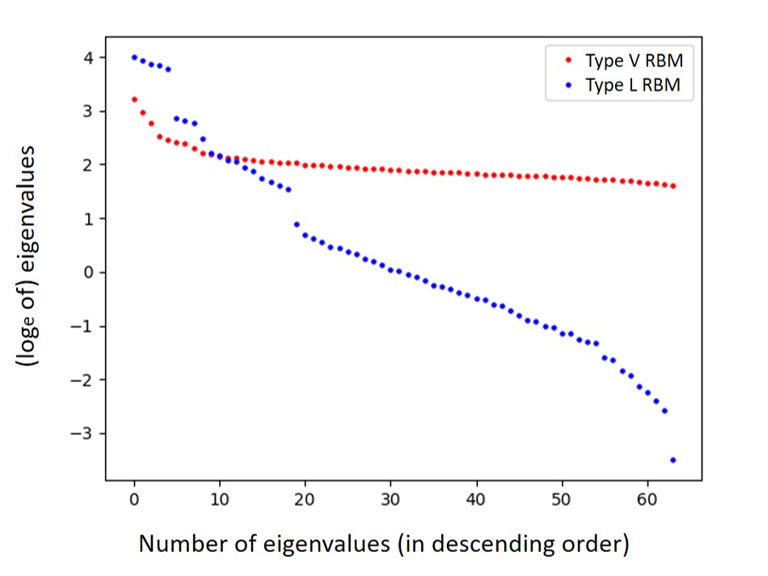}
\caption{\small 
Eigenvalues of $WW^T$ for type V RBM (red, a smooth line) and type L RBM (blue, a steplike line).
Both RBMs have 64 neurons in the hidden layer.}
\label{fig10_0}
\end{center}
\end{figure}

\begin{figure}[h]
\begin{center}
\includegraphics[width=8cm, height=6cm]{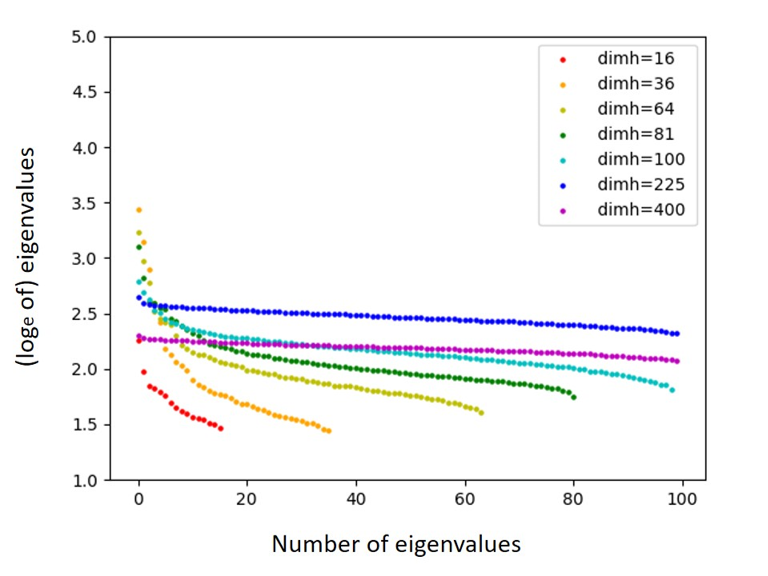}
\includegraphics[width=8cm, height=6cm]{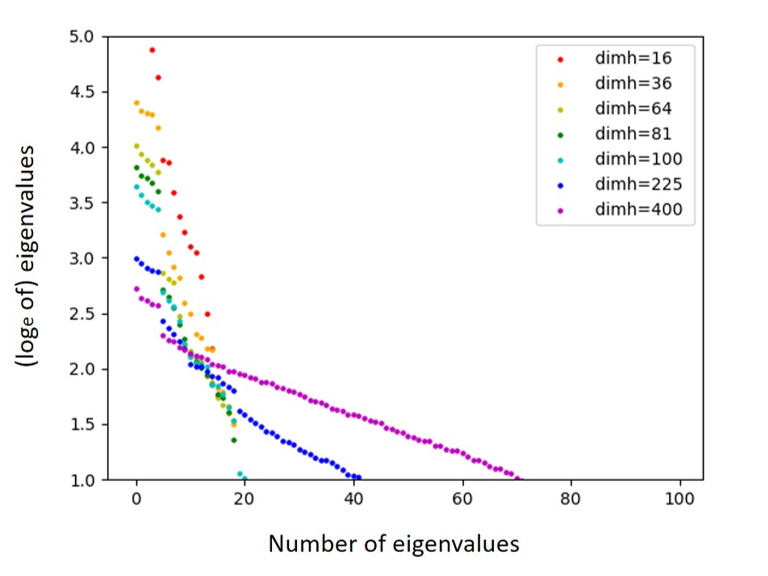}
\caption{\small Eigenvalues of $WW^T$ for type V RBM  (left) and type L RBM (right).
The legend shows the number of hidden neurons $N_h$.}
\label{fig10}
\end{center}
\end{figure}

Finally we study the eigenvalue spectrum $\lambda_a$ of the matrix $WW^T$.
Figs.\,\ref{fig10_0} and \ref{fig10} show the eigenvalues  in the descending order.
In Fig.\,\ref{fig10_0}, 
the red dots (a smooth line) show the eigenvalues of the type V RBM trained by configurations at 
all the temperatures ($T=0,0.25,\ldots,6$), 
while the blue dots (a steplike line) are the eigenvalues of the type L RBM (only $T=0$).

These are obviously different. 
For the type L RBM, only several eigenvalues are especially large,
and the rest are apparently smaller. 
On the other hand, for the type V RBM, the eigenvalues decrease gradually and there are no jumps or big
distinctions between larger and smaller eigenvalues.
The behavior indicates that, in the type L RBM, 
the weight matrix holds only small relevant information
and only a small number of neurons is sufficient in the hidden layer.
In the type V RBM, however, since it is trained by configurations at various different temperatures,
all the eigenvectors are equally utilized to represent relevant features of spin configurations at various temperatures. 
Namely, 
in order to learn  features of a wide range of temperatures,
larger number of neurons in the hidden layer are necessary\footnote{
However, 
too many hidden neurons ($N_h>N_v$) are not appropriate because lots of irrelevant features 
are acquired.}. 
Using such larger degrees of freedom,
the weight matrix has learned  configurations with various characteristic scales 
at various temperatures so that
the RBM can grasp rich properties of these configurations.

The difference of the eigenvalues  between type V and type L is also phrased 
that type V has a scale invariant eigenvalue spectrum\footnote{
Exactly speaking, it is not completely scale invariant, 
but compared to type L, there is at least no jump between larger and smaller
eigenvalues.}. 
In contrast, the eigenvalues of the type L RBM are separated into distinct regions
in which the corresponding eigenvalues might represent 
 features with different scales.
It might be related to our previous numerical results, shown in 
 Figs.\,\ref{fig2}, \ref{fig2_6} and \ref{fig3}, that type V RBM generates a flow toward the critical point
{where the configurations have scale invariance}.  

Finally in Fig.\,\ref{fig10} we show
 differences in eigenvalue spectrum between RBMs with different numbers of hidden neurons $N_h$.
As shown in the left panel of Fig.\,\ref{fig10}, in the type V RBM with $N_h>N_v=100$,
most eigenvalues  have similar values. 
In contrast, for a smaller $N_h$,  large and small eigenvalues 
are very different and the spectrum has a hierarchical structure.
The type L RBM  shows similar behaviors as shown in the right panel of Fig.\,\ref{fig10}.  
It might indicate that the RBM with larger $N_h$ ($>N_v$)  
has learned too much details of the input configurations and the most relevant
features are weakened.
In other words, it has learned too much irrelevant features which are especially specific
to configurations at higher temperature.
It may explain our numerical results shown in Fig.\,\ref{fig4}, 
which is apparently different from Fig.\,\ref{fig2},
that the flow first approaches the critical point but passes there and eventually goes
away to higher temperature.
This view is consistent with the discussion at the end of Sec.\,\ref{sec:SVD}
and  supports the second conjecture in Sec.\,\ref{sec:res_rbm}.

To summarize, we find that the
type V RBM with smaller $N_h$ than $N_v$ can adequately learn 
configurations at  wide range of temperatures,
without learning too much features at  higher temperature.
All the neurons in the hidden layer are efficiently used to represent
features of various temperatures in a scale invariant way as seen 
in the eigenvalue spectrum. 
As a result, after many iterations of the RBM reconstruction, 
initial configurations are transformed into the configurations around the critical temperature $T_c$. 
Thus the RBM flow has similarity with the RG flow, but
the RBM flow is  in the opposite direction to the conventional RG flow and 
a naive analogy does not hold. 




\section{Discussions}

In this paper, in order to see
what the restricted Boltzmann machine (RBM)  learns in the process of training,
we investigated  flow of configurations
generated by the weight matrices of the RBM.
In particular, we studied the Ising model and found
 that, if the RBM is trained
by spin configurations at various different temperatures (we call it type V RBM), 
the temperature of an initial configuration 
  flows towards the critical point $T=T_c$ where the system becomes scale invariant.
The result suggests that the configurations at $T=T_c$ are {\it attractors} of the RBM flow.
In order to understand the numerical results of the RBM flows
and to find a clue of what the machine has learned, 
we  explored  properties of the weight matrix $W_{ia}$ , especially
those of the product $WW^T$,  by looking at the eigenvalue spectrum.

There are still many unsolved issues left for future investigations.
If we admit that an eigenvector represents a ``feature" that the RBM has learned,
the magnitude of the corresponding eigenvalue is an indicator of how much {influence}
the feature affects, and
reminiscent of  the critical exponents of relevant operators in the renormalization group (RG).
It will be interesting to pursue this analogy further and to extract more information
from the eigenvalue spectrum and to connect with the behaviors of the RBM flow.
The RBM flow gives  important clues of what the machine has learned. 
If an RBM is trained by configurations at a single specific temperature, it generates a 
flow toward that temperature. This  confirms a hypothesis that the unsupervised RBM indeed extracted 
relevant features of the configurations and the flow is consequently attracted to the configurations
with these relevant features. 
The RBM flow is a stochastic process with a random noise, and
  should be described by the Langevin (or Fokker Planck) equation, whose drift term is given by
  the relevant features. 
We want to come back to this problem in future investigations.

In the present paper, we picked up the Ising model, the simplest statistical model 
of the second order phase transition and  found that 
the critical point is an attractor of the RBM flow. 
Then we wonder what happens in the case of the first order phase transition.
A simple example is the Blume-Capel model on a two dimensional lattice.
The Hamiltonian is given by $\beta H=-J\sum_{ij} s_i s_j + \Delta \sum_i s_i^2$ where 
 $s_i=\pm 1, 0$ is a spin  at site  $i$. 
The model undergoes the first order phase transition that separates the parameter space 
$(J,\Delta)$, 
and the second order phase transition at the tricritical point 
$(J_c, \Delta_c)$.
 If we train an RBM by configurations
at various different parameters (such as the type V RBM  of the Ising model), 
the flow of parameters will be attracted to the tricritical point. 
On the other hand, 
if we use only a restricted set of configurations for training, 
e.g.  various $J$ with a fixed $\Delta(<\Delta_c)$, 
where is the RBM flow attracted? 
It is not certain whether it is attracted to the phase boundary of the first order phase transition
or to the tricritical point. 
It is under investigations and
we want to report the numerical results in future publications. 

Finally we would like to comment on a possible relation between  structures
of RBM flows and how we recognize the world around us.
Our finding is that there is an attractor of the RBM flow
which characterizes the relevant feature that machine has learned before.
We human beings also meet similar phenomena, namely we can 
recognize more easily and comfortably what we have already learned many times than
what we first experience. Also what we think beautiful is not what we 
experience first but usually a combination of 
what  we have already experienced before. 
A good example will be {looking at} abstract paintings or {tasting} bitter coffee. 
It suggests that attractors are constructed in the process of learnings
and we {would} feel comfortable when input data are close to the attractors 
of the neural network in our brain. 
For verification of the conjecture, it {must be} amusing to train a RBM by inputting varieties of human faces
and generate the RBM flow of a human face to see if there is an attractor face. 
The attractor {might} provide a standard for beauty.
In this way, we can guess that
attractors of the RBM flows in the NN of our brain may 
control our value judgments.  

\subsection*{Acknowledgments}

We would like to thank participants of I-URIC frontier colloquium 2017,
especially Shunichi Amari, Taro Toyoizumi and Shinsuke Koyama
for fruitful discussions.
We also thank Masato Taki for his intensive lectures on machine learning at KEK.
This work of S.I. and S.S. is supported in part by Grants-in-Aid for Scientific Research (No.\,16K05329 and No.\,16K17711, respectively) from the Japan Society for the Promotion of Science (JSPS).

\end{document}